\documentclass{article}[12pt]
\usepackage{graphicx}
\usepackage{setspace}
\usepackage{mathrsfs}
\usepackage{float}
\usepackage{multicol}
\usepackage{amsmath,amsfonts,amssymb,amsthm}
\usepackage{cite}
\usepackage{mathrsfs}
\usepackage{hyperref}
\usepackage{indentfirst}
\usepackage[toc,page]{appendix}
\usepackage{mathtools}
\usepackage{tikz}
\usepackage{pgfplots}
\pgfplotsset{compat=1.18}
\numberwithin{equation}{section}

\newtheorem{theorem}{Theorem}[section]

\newtheorem{remark}{Remark}[section]
\newcommand\be{\begin{equation}}
\newcommand\ee{\end{equation}}
\newcommand\ber{\begin{eqnarray}}
\newcommand\eer{\end{eqnarray}}
\newcommand\berr{\begin{eqnarray*}}
\newcommand\eerr{\end{eqnarray*}}
\allowdisplaybreaks
\begin{document}
\title{ \bf\Large Finite Energy, Event Horizons, and a Universal Bound on the Singularity Order in Born--Infeld-type Black Holes}
\author{\bf{Tengyang Liu}$^{a,}$\footnote{Email address: tyl619@henu.edu.cn}\\[2mm]{\it\small $^a$College of Mathematics and Statistics, Henan University}\\[2mm]{\it\small Kaifeng 475001, P.R.China}}
\date{}
\maketitle
\begin{abstract}
We investigate an interpolated Maxwell--Born--Infeld theory bridging the Maxwell and Born--Infeld limits. In the electrostatic sector the total energy is finite, and remains bounded above even as the theory approaches the Maxwell limit. We construct the electrically charged black hole, whose black hole solution interpolates monotonically between the classical Born--Infeld and Reissner--Nordström solutions; singularity can be relegated, independently of the weight parameters. For general interpolated theories with a truncation threshold, we analyze the existence and multiplicity of event horizons and classify the horizon configurations, and for general Born--Infeld-type nonlinear structures we prove a universal lower bound on the order of the Kretschmann divergence, thereby excluding regular electric black holes.

\medskip


\end{abstract}

\section{Introduction}

The classical Maxwell theory of electromagnetism, despite its extraordinary success in describing a vast range of physical phenomena, suffers from a well-known shortcoming: the electrostatic field energy of a point charge diverges, leading to an infinite self-energy. Such a divergence indicates that the linear Maxwell theory ceases to be reliable at short distances, and it has prompted many attempts to build nonlinear generalizations that tame the singular behaviour without spoiling the Maxwell limit on large scales. The Maxwell theory is not unique in this respect, however: it places electric and magnetic point charges on an equal footing and assigns an infinite energy to each of them. A magnetic point charge, or monopole, has never been observed, although it was advocated by Curie\cite{Curie} as early as 1894 and formulated mathematically by Dirac in 1931\cite{Dirac}; its absence, however, is not implied by the Maxwell action itself but must be imposed by hand.

A landmark achievement in this direction is the classical Born--Infeld theory \cite{B1,B2,BI1,BI2}, which introduces a maximal electric field strength $1/\sqrt{\beta}$ through a square-root Lagrangian
\begin{align}
    \mathcal{L}_{\rm BI}(s)=\frac{1}{\beta}\left(1-\sqrt{1-2\beta s}\right).
\end{align}
Here, $s$ may denote not only the Maxwell action density $\mathcal{L}_{\rm M}$, but also the Born--Infeld action density including the second Born--Infeld term \cite{Y-book}, given by
\begin{align}
    s:=\mathcal{L}_{\rm M}+\frac{\kappa^2}{32}\left(F_{\mu\nu}\tilde{F}^{\mu\nu}\right)^2=-\frac{1}{4}F_{\mu\nu}F^{\mu\nu}+\frac{\kappa^2}{32}\left(F_{\mu\nu}\tilde{F}^{\mu\nu}\right)^2,\quad \tilde{F}^{\mu\nu}=\frac12\epsilon^{\mu\nu\alpha\xi}F_{\alpha\xi},\quad \mu,\nu=0,1,2,3,\quad\label{1.2}
\end{align}
where $F_{\mu\nu}$ is the electromagnetic field strength and $\kappa\ge0$ is a free coupling parameter.

On the one hand, the classical Born--Infeld theory cures the electric divergence and yields a finite total energy for an electric point charge, from which \cite{LY1} estimated an effective electron radius $r_0\approx2.06\,r_{\rm e}$, where $r_{\rm e}\approx1.1067\times10^{-13}\,{\rm cm}$ is the classical electron radius, corresponding to a Born parameter $b\approx9.24\times10^{15}\,{\rm cm}^{-1/2}{\rm g}^{1/2}{\rm s}^{-1}$. Two features of this outcome deserve emphasis. First, the estimated radius is roughly twice the classical electron radius, which is larger than what one would expect for a point-like particle; a theory predicting a somewhat smaller effective radius would be closer to observation. Second, and more fundamentally, the classical Born--Infeld theory assigns a finite energy to a magnetic point charge as well, so that it accommodates a monopole just as it accommodates an electric charge, thereby exhibiting an electromagnetic symmetry that the Maxwell theory also possesses, albeit with the roles of finite and infinite energies interchanged. Since monopoles have never been detected, a nonlinear theory that retains the electric regularity of Born--Infeld while excluding finite-energy monopoles would be physically more satisfactory. On the other hand, the coupling of the classical Born--Infeld theory to general relativity has been extensively studied, and the resulting black hole solutions have long served as a prototype of nonlinear electrodynamics interacting with gravity. Charged black hole solutions were obtained early on \cite{Hoffmann,PellicerTorrence}, and subsequently extended to backgrounds with a cosmological constant and to asymptotically (A)dS spacetimes \cite{Dey,Cai}, to rotating and extremal configurations \cite{RotatingBI}, and to Eddington-inspired Born--Infeld gravity, where geonic black holes and remnants arise \cite{OlmoGeonic}. These studies show that the nonlinear structure of the Born–Infeld Lagrangian has clear consequences for the black hole. In particular, it affects the horizon structure, the asymptotic behaviour and the singularity. Based on what has been stated above, people begin to search for new nonlinear theories, namely the Born--Infeld-type nonlinear electrodynamics, in order to answer or further explore the aforementioned topics.

Recent years have seen extensive work on Born–Infeld-type theories. Their physical properties, such as energy finiteness, causality and the existence of charged black hole solutions, differ substantially from model to model\cite{BretOn2007,Jimenez2017,RussoTownsend2024}. For polynomial models of the form $\mathcal{L}=\sum_{n\ge1}\zeta_n s^n,\zeta_1=1$, the electromagnetic asymmetry is known to hold \cite{yang2022electromagnetic}: an electric point charge can acquire finite energy, whereas finite-energy monopoles are excluded. The Stone--Weierstrass theorem \cite{Stone1948,Yosida1995} shows that on a compact region every continuous nonlinear Lagrangian can be uniformly approximated by such polynomial models, so polynomial models constitute a sufficiently covering class of exactly analyzable benchmarks; this, however, does not automatically guarantee that the monopole-exclusion property survives the approximation limit, nor does it justify the assertion that every Born--Infeld-type model excludes monopoles. In other words, the polynomial model results establish monopole exclusion within a broad class of tractable models, while whether more general Born--Infeld-type theories share this property requires model-by-model verification. The classical Born--Infeld Lagrangian, with its square-root structure, is precisely the exceptional case in which the symmetry survives \cite{Ydyonic}. The question therefore arises: can one construct a theory that retains the electric regularity of Born--Infeld, reduces its effective electron radius, and excludes finite-energy monopoles at the same time?

In our previous work \cite{LY1}, we answered these questions by introducing an interpolated model in which the nonlinear Born--Infeld correction acts as a controlled perturbation of the Maxwell theory. The Lagrangian is constructed as a convex combination of the Maxwell and classical Born--Infeld Lagrangians,
\begin{align}
    \mathcal{L}_{\rm MBI}=\kappa_1 \mathcal{L}_{\rm M}+\frac{\kappa_2}{\beta}\left(1-\sqrt{1-2\beta s}\right),\quad \kappa_1+\kappa_2=1,\quad \kappa_1\in[0,1),\ \kappa_2\in(0,1], \quad\mathcal{L}_{\rm M}=-\frac{1}{4}F_{\mu\nu}F^{\mu\nu},\label{1.4}
\end{align}
where $s$ is given by \eqref{1.2} and the electromagnetic field $F_{\mu\nu}=\partial_\mu A_\nu-\partial_\nu A_\mu$ is induced by a real-valued gauge field $A_{\mu}$. This parametrization has a clear physical reading. The Maxwell and Born–Infeld terms are blended with weights $\kappa_1$ and $\kappa_2$, respectively. It satisfies the standard normalization conditions for any choice of $\kappa_1,\kappa_2$, and smoothly interpolates between the two limiting cases: $\kappa_1=0$ recovers the classical Born--Infeld theory, while $\kappa_1\to1$ ($\kappa_2\to0$) approaches the Maxwell theory in a controlled manner. Unlike the conventional approach of constructing generalized Born--Infeld theories by imposing only the normalization conditions--which often leads to a vast and largely arbitrary landscape of models--our interpolation scheme provides a systematic and physically transparent framework for studying how the nonlinear Born--Infeld correction is gradually switched off, and how the physical properties of the system evolve from the Born--Infeld regime to the Maxwell regime. The combination is convex, so the resulting theory inherits properties from both limiting theories. The relative weight of the nonlinear correction is controlled by the single free parameter $\kappa_1$ (or equivalently $\kappa_2$). This makes $\kappa_1$ a natural dial for probing the transition from nonlinear to linear electrodynamics.

In \cite{LY1} it was shown that the interpolated theory does achieve the desired asymmetry: it admits a finite-energy electric point charge, as in the classical Born--Infeld theory, but no finite-energy monopole or dyon, since the Maxwell term contributes a divergence  to the energy of any magnetic configuration. They also found that the normalized energy falls monotonically as $\kappa_2$ increases. A larger Maxwell weight therefore gives a higher energy and, through the energy–radius relation $E/(4\pi)=(q^2/a)H(\kappa_1,\kappa_2)$, a larger effective electron radius. The interpolated theory therefore raises, rather than resolves, the question of the effective electron radius: the Maxwell weight needed to suppress monopoles simultaneously inflates the radius.

The analysis of \cite{LY1} was carried out entirely in Minkowski spacetime, and its energy integral was evaluated by interchanging the limit $\kappa_2\to0$ with the integration over $\eta$. The present paper first revisits this energy integral in the Maxwell limit, where the naive interchange produces the artifact $1/2$ and misses an infinitesimally thin boundary layer near $\eta=1$; we show that the correct sharp upper bound of the electric energy is $4/3$ (Section \ref{s3.1}). We then couple the theory to gravity, where it yields charged black holes, and take up two questions that the flat-spacetime analysis could not reach. First, how does the interpolated black hole interpolate between the classical Born--Infeld and Reissner--Nordstr\"om solutions as the weight $\kappa_1$ varies, and how does its curvature singularity behave? Second, are the magnetic and dyonic configurations, already ruled out on energetic grounds in \cite{LY1}, still excluded once gravity is included, and does gravity change the way this energetic exclusion works? Beyond the interpolated model itself, we further extend the analysis to general interpolated Born--Infeld-type theories of the form $\mathcal{L}(s)=\kappa_1\mathcal{L}_{\rm M}+\frac{\kappa_2}{\beta}U(2\beta s)$, where the nonlinear structure $U$ is required only to satisfy the normalization conditions $U(0)=0$, $U'(0)=1$ and the convexity conditions $U'>0$, $U''>0$ of the Pleba\'nski class\cite{Plebanski1968,Schellstede2016,Russo2024}, the admissible class being further enlarged by polynomial approximation via the Stone--Weierstrass theorem, and may either be defined on the whole range of $s$ attained by the field configurations under consideration or possess a truncation threshold $s<1/(2\beta)$. For this general class, the Kretschmann curvature is shown to exhibit the Schwarzschild-type $r^{-6}$ divergence near the origin in the generic case, the relegation is shown to stop at the universal bound $r^{-4}$  (equivalently, the divergence order cannot be made weaker than $r^{-4}$) on the strength of the Kretschmann divergence, which is independent of the weight parameters $\kappa_1,\kappa_2$ (though the precise exponent may depend on the growth class of $U$), so that regular electric black holes are excluded in a model-independent way, and the event horizon configurations are completely classified.

This paper is organized as follows. In Section \ref{s2} we present the general setup of Born--Infeld-type theories and derive the field equations and the energy-momentum tensor. In Section \ref{s3} we construct the electrically, magnetically and dyonically charged black hole solutions, analyze their asymptotic behavior, total energy and Kretschmann curvature, and examine the energy conditions in the electric, magnetic and dyonic sectors. In Section \ref{s4} we investigate the existence and multiplicity of event horizons for the interpolated electric black holes and the influence of the interpolation and Born parameters on the horizon radii. In Section \ref{s5} we extend the analysis to general interpolated Born--Infeld-type theories, establishing the lower bound on the order of the Kretschmann divergence and the classification of horizon configurations. Finally, Section \ref{s6} contains our conclusions and outlook.

\section{Formalism}\label{s2}

We work first on the Minkowski spacetime with metric
$(\xi_{\mu\nu})=\mathrm{diag}(1,-1,-1,-1)$ and coordinates
$x^\mu=(t,\mathbf x)$. Let $A_\mu$ be a real-valued gauge potential and
\begin{align}
    F_{\mu\nu}=\partial_\mu A_\nu-\partial_\nu A_\mu
    \label{2.1}
\end{align}
the electromagnetic field strength. Following the sign convention of \cite{Ryder,Huang}, we define the electric and magnetic fields by
\begin{align}
    E^i=-F^{0i},\quad B^i=-\frac12\epsilon^{ijl}F_{jl},\quad i,j,l=1,2,3,
    \label{2.2}
\end{align}
so that the matrix forms of $F^{\mu\nu}$ and its dual
$\tilde F^{\mu\nu}=\frac12\epsilon^{\mu\nu\alpha\gamma}F_{\alpha\gamma}$ read
\begin{align}
    (F^{\mu\nu})=
    \begin{pmatrix}
        0 & -E^1 & -E^2 & -E^3\\
        E^1 & 0 & -B^3 & B^2\\
        E^2 & B^3 & 0 & -B^1\\
        E^3 & -B^2 & B^1 & 0
    \end{pmatrix},
    \label{2.3}\quad 
    (\tilde F^{\mu\nu})=
    \begin{pmatrix}
        0 & -B^1 & -B^2 & -B^3\\
        B^1 & 0 & E^3 & -E^2\\
        B^2 & -E^3 & 0 & E^1\\
        B^3 & E^2 & -E^1 & 0
    \end{pmatrix},
\end{align}
where $\epsilon^{0123}=1$ in Minkowski spacetime. The dual tensor
realizes the electromagnetic duality $\mathbf E\mapsto\mathbf B$,
$\mathbf B\mapsto-\mathbf E$, and the Bianchi identity
$\partial_\mu\tilde F^{\mu\nu}=0$ follows from \eqref{2.1}.
From \eqref{2.3} we compute the two fundamental invariants
\begin{align}
    F_{\mu\nu}F^{\mu\nu}&=-2\left(\mathbf E^2-\mathbf B^2\right),
    \label{2.5}\\
    F_{\mu\nu}\tilde F^{\mu\nu}&=-4\,\mathbf E\cdot\mathbf B.
    \label{2.6}
\end{align}
The generalized Born--Infeld action density is then defined by\cite{Y-book,Y-dyonic2022}
\begin{align}
    s:=-\frac14F_{\mu\nu}F^{\mu\nu}+\frac{\kappa^2}{32}\left(F_{\mu\nu}\tilde F^{\mu\nu}\right)^2
    =\frac12\left(\mathbf E^2-\mathbf B^2\right)+\frac{\kappa^2}{2}\left(\mathbf E\cdot\mathbf B\right)^2,
    \label{2.7}
\end{align}
where $\kappa\ge0$ is a free coupling parameter. In particular, $s$ reduces
to the Maxwell action density
$\mathcal L_{\rm M}=\frac12(\mathbf E^2-\mathbf B^2)$ when $\kappa=0$, and
the second term in \eqref{2.7} is the second Born--Infeld action density.

For an arbitrary Born--Infeld type lagrangian $\mathcal L(s)$, the Euler--Lagrange equation
associated with $A_\mu$ is
\begin{align}
    \partial_\mu P^{\mu\nu}=0,\quad
    P^{\mu\nu}=\mathcal{L}'(s)\left(F^{\mu\nu}-\frac{\kappa^2}{4}\left(F_{\alpha\gamma}\tilde F^{\alpha\gamma}\right)\tilde F^{\mu\nu}\right),
    \label{2.8}
\end{align}
The electric displacement and magnetic
intensity fields are
\begin{align}
    D^i=-P^{0i},\quad H^i=-\frac12\epsilon^{ijk}P_{jk},
    \label{2.9}
\end{align}
so that \eqref{2.8} takes the vacuum Maxwell form
\begin{align}
\partial_t\mathbf B+\nabla\times\mathbf E=\mathbf 0,\quad
\nabla\cdot\mathbf B=0,\quad
-\partial_t\mathbf D+\nabla\times\mathbf H=\mathbf 0,\quad
\nabla\cdot\mathbf D=0, 
\end{align}
while the constitutive relations are
\begin{align}
    \mathbf D=\mathcal{L}'(s)\left(\mathbf E+\kappa^2(\mathbf E\cdot\mathbf B)\mathbf B\right),\quad
    \mathbf H=\mathcal{L}'(s)\left(\mathbf B-\kappa^2(\mathbf E\cdot\mathbf B)\mathbf E\right).
    \label{2.10}
\end{align}
The energy-momentum tensor is
\begin{align}
    T_{\mu\nu}
    &=-\mathcal{L}'(s)\left(F_{\mu\alpha}\xi^{\alpha\gamma}F_{\nu\gamma}
    -\frac{\kappa^2}{4}\left(F_{\alpha\gamma}\tilde F^{\alpha\gamma}\right)F_{\mu\alpha}\xi^{\alpha\gamma}\tilde F_{\nu\gamma}\right)-\xi_{\mu\nu}\mathcal{L}(s),
    \label{2.11}
\end{align}
and the Hamiltonian energy density reads
\begin{align}
    \mathcal H=T_{00}=\mathcal{L}'(s)\left(\mathbf E^2+\kappa^2(\mathbf E\cdot\mathbf B)^2\right)-\mathcal{L}(s).
    \label{2.12}
\end{align}

We now couple the theory $\mathcal{L}(s)$ to gravity and specialize to a static, spherically symmetric spacetime with the line element
\begin{align}
    \mathrm ds^2=A(r)\mathrm dt^2-\frac{\mathrm dr^2}{A(r)}-r^2\left(\mathrm d\theta^2+\sin^2\theta\,\mathrm d\phi^2\right),
    \label{2.14}
\end{align}
in the ordered coordinates $(x^\mu)=(t,r,\theta,\phi)$. In this background the metric determinant is $\sqrt{-g}=r^2\sin\theta$, and the dual tensor $\tilde F^{\mu\nu}$ is replaced by
\begin{align}
    \tilde F^{\mu\nu}=\frac12\epsilon^{\mu\nu\alpha\gamma}F_{\alpha\gamma},\quad
    \epsilon^{0123}=\frac{1}{\sqrt{-g}}=\frac{1}{r^2\sin\theta}.
    \label{2.15}
\end{align}
For a dyonic configuration carrying both electric and magnetic charges, we take the field strength in the form
\begin{align}
    (F_{\mu\nu})=
    \begin{pmatrix}
        0 & E^r & 0 & 0\\
        -E^r & 0 & 0 & 0\\
        0 & 0 & 0 & -r^4\sin^2\theta\,B^r\\
        0 & 0 & r^4\sin^2\theta\,B^r & 0
    \end{pmatrix},
    \label{2.16}
\end{align}
where $E^r$ and $B^r$ are the radial electric and magnetic fields. It is
convenient to factor out the angular dependence by setting
\begin{align}
    B^r=\frac{B(r)}{\sin\theta},
    \label{2.17}
\end{align}
with $B(r)$ a radial function to be determined. Then \eqref{2.15}--\eqref{2.32}
give
\begin{align}
    (\tilde F^{\mu\nu})=
    \begin{pmatrix}
        0 & -r^2B(r) & 0 & 0\\
        r^2B(r) & 0 & 0 & 0\\
        0 & 0 & 0 & \dfrac{E^r}{r^2\sin\theta}\\[4pt]
        0 & 0 & -\dfrac{E^r}{r^2\sin\theta} & 0
    \end{pmatrix},
    \label{2.18}
\end{align}
and the two invariants become
\begin{align}
    F_{\mu\nu}F^{\mu\nu}=-2(E^r)^2+2r^4B(r)^2,\quad 
    F_{\mu\nu}\tilde F^{\mu\nu}=-4r^2E^rB(r).\label{2.19}
\end{align}
Consequently, the generalized Born--Infeld action density \eqref{2.7} reduces to
\begin{align}
    s=\frac12\left((E^r)^2-r^4B(r)^2\right)+\frac{\kappa^2}{2}r^4\left(E^rB(r)\right)^2.
    \label{2.21}
\end{align}
The Bianchi identity $\partial_\mu(\sqrt{-g}\,\tilde F^{\mu\nu})=0$ with
\eqref{2.18} yields
\begin{align}
    \partial_r\left(r^4\sin\theta\,B^r\right)=0,
    \label{2.22}
\end{align}
so that $r^4B(r)=g$ is a constant, which we identify with the magnetic
charge:
\begin{align}
    B(r)=\frac{g}{r^4}.
    \label{2.23}
\end{align}
Substituting \eqref{2.23} into \eqref{2.19}--\eqref{2.21}, we obtain
\begin{align}
    F_{\mu\nu}F^{\mu\nu}&=-2(E^r)^2+\frac{2g^2}{r^4},
    \label{2.24}\\
    F_{\mu\nu}\tilde F^{\mu\nu}&=-\frac{4gE^r}{r^2},
    \label{2.25}\\
    s&=\frac12\left((E^r)^2-\frac{g^2}{r^4}\right)+\frac{\kappa^2g^2}{2r^4}(E^r)^2.
    \label{2.26}
\end{align}
Equations \eqref{2.24}--\eqref{2.26} are the curved-spacetime counterparts
of \eqref{2.5}--\eqref{2.7}, and they coincide with the Minkowski expressions
upon replacing $\mathbf E^2\to(E^r)^2$, $\mathbf B^2\to g^2/r^4$,
$\mathbf E\cdot\mathbf B\to gE^r/r^2$.

The generalized Maxwell equations in the curved background read
\begin{align}
    \frac{1}{\sqrt{-g}}\partial_\mu\left(\sqrt{-g}\,P^{\mu\nu}\right)=0,
    \label{2.27}
\end{align}
with $P^{\mu\nu}$ given by \eqref{2.8}. Using \eqref{2.18}, \eqref{2.23},
and \eqref{2.25}, the non-vanishing components of $P^{\mu\nu}$ are
\begin{align}
    P^{0r}&=-P^{r0}
    =\left(\kappa_1+\frac{\kappa_2}{\sqrt{1-2\beta s}}\right)
    \left(1+\frac{\kappa^2g^2}{r^4}\right)E^r,
    \label{2.28}\\
    P^{\theta\phi}&=-P^{\phi\theta}
    =\left(\kappa_1+\frac{\kappa_2}{\sqrt{1-2\beta s}}\right)
    \left(1-\kappa^2(E^r)^2\right)\frac{g}{r^4\sin\theta}.
    \label{2.29}
\end{align}
Equation \eqref{2.27} then reduces to
\begin{align}
    \partial_r\left(r^2P^{0r}\right)=0,\quad
    \partial_\theta\left(\sin\theta\,P^{\theta\phi}\right)=0.
    \label{2.30}
\end{align}
Integrating the first equation gives
\begin{align}
    r^2P^{0r}=q,
    \label{2.31}
\end{align}
with $q$ the electric charge, and combining \eqref{2.28} and \eqref{2.31}
yields the constitutive equation determining $E^r$:
\begin{align}
    \frac{q}{r^2}
    =\left(\kappa_1+\frac{\kappa_2}{\sqrt{1-2\beta s}}\right)
    \left(1+\frac{\kappa^2g^2}{r^4}\right)E^r,
    \label{2.32}
\end{align}
with $s$ given by \eqref{2.26}. The second equation in \eqref{2.30} is
automatically satisfied by \eqref{2.29} and \eqref{2.23}. Thus the
electromagnetic sector is fully encoded in the single implicit equation
\eqref{2.32} for $E^r$, with $g$ entering as the magnetic charge parameter.

The energy-momentum tensor in the curved background is
\begin{align}
    T_{\mu\nu}
    &=2\frac{\partial\mathcal L_{\rm MBI}(s)}{\partial g^{\mu\nu}}-g_{\mu\nu}\mathcal L_{\rm MBI}(s) \notag\\
    &=-\mathcal{L}_{\rm MBI}'(s)\left(F_{\mu\alpha}g^{\alpha\gamma}F_{\nu\gamma}-\frac{\kappa^2}{4}\left(F_{\alpha\gamma}\tilde F^{\alpha\gamma}\right)F_{\mu\alpha}g^{\alpha\gamma}\tilde F_{\nu\gamma}\right)-g_{\mu\nu}\mathcal{L}_{\rm MBI}(s).\label{2.33}
\end{align}
For our interpolated Maxwell--Born--Infeld theory \eqref{1.4}, using \eqref{2.16}--\eqref{2.26}, the non-vanishing components are
\begin{align}
    T_{00}&=A(r)\mathcal H^{q,g}_{\rm MBI},
    \label{2.34}\\
    T_{11}&=-A(r)^{-1}\mathcal H^{q,g}_{\rm MBI},
    \label{2.35}\\
    T_{22}&=r^6\left(\kappa_1+\frac{\kappa_2}{\sqrt{1-2\beta s}}\right)\left(\frac{g^2}{r^8}-\kappa^2\frac{(E^r)^2g^2}{r^8}\right)+r^2\left[\kappa_1s+\frac{\kappa_2}{\beta}\left(1-\sqrt{1-2\beta s}\right)\right],
    \label{2.36}\\
    T_{33}&=\sin^2\theta\,T_{22},
    \label{2.37}
\end{align}
where the Hamiltonian energy density is
\begin{align}
    \mathcal H^{q,g}_{\rm MBI}
    &=\frac{\kappa_1}{2}\left((E^r)^2+\frac{g^2}{r^4}\right)
    +\frac{\kappa_2}{\sqrt{1-2\beta s}}\left(\frac{g^2}{r^4}+\frac1\beta\right)-\frac{\kappa_2}{\beta}\notag\\
    &\quad+\frac{\kappa_1\kappa^2g^2}{2r^4}(E^r)^2
    +\frac{\kappa_2\kappa^2g^2}{r^4\sqrt{1-2\beta s}}(E^r)^2.
    \label{2.38}
\end{align}
Equation \eqref{2.38} is the curved-spacetime version of \eqref{2.12} on the premise of \eqref{1.4}; in the purely electric case $g=0$ it reduces to \eqref{HMBI}, and in the purely magnetic case $E^r=0$ it reduces to the magnetic energy density used in Section \ref{s3}. The trace of the energy-momentum tensor is
\begin{align}
    T=g^{\mu\nu}T_{\mu\nu}
=\frac{2\kappa_1g^2}{r^4}+\frac{2\kappa_2}{\sqrt{1-2\beta s}}\left((E^r)^2-\frac{g^2}{r^4}+2\frac{\kappa^2g^2}{r^4}(E^r)^2\right)-4\mathcal L_{\rm MBI}(s).
    \label{2.39}
\end{align}

The Einstein equations are
\begin{align}
    R_{\mu\nu}=-8\pi G\left(T_{\mu\nu}-\frac12g_{\mu\nu}T\right),
    \label{2.40}
\end{align}
and for the metric \eqref{2.14} the non-vanishing Ricci components are
\begin{align}
    R_{00}=-\frac{A}{2r^2}(r^2A')',\quad
    R_{11}=\frac{1}{2r^2A}(r^2A')',\quad
    R_{22}=(rA)'-1,\quad
    R_{33}=\sin^2\theta\,R_{22}.
    \label{2.41}
\end{align}
Substituting \eqref{2.34}--\eqref{2.39} and \eqref{2.41} into \eqref{2.40},
we have the time-time and $rr$ components
\begin{align}
    -\frac{A}{2r^2}(r^2A')'=4\pi GA\left(\mathcal H-\frac12T\right),\quad \frac{1}{2r^2A}(r^2A')'=4\pi GA^{-1}\left(\mathcal H-\frac12T\right).
\end{align}
These two equations are consistent and reduce to the first-order equation
\begin{align}
    (rA)'=1-8\pi Gr^2\mathcal H.
    \label{2.44}
\end{align}
Integrating \eqref{2.44} yields
\begin{align}
    A(r)=1-\frac{2GM}{r}+\frac{8\pi G}{r}\int_r^\infty\mathcal H(\rho)\rho^2\mathrm d\rho,
    \label{2.45}
\end{align}
where $M>0$. The total electromagnetic energy is
\begin{align}
    E=4\pi\int_0^\infty\mathcal H(r)r^2\mathrm dr.
    \label{2.46}
\end{align}
By \eqref{2.46}, we can rewrite \eqref{2.45} as
\begin{align}
    A(r)=1-\frac{2G(M-E)}{r}-\frac{8\pi G}{r}\int_0^r\mathcal H(\rho)\rho^2\mathrm d\rho.
    \label{2.47}
\end{align}

\section{Black hole}\label{s3}

In \cite{LY1}, we have reached the conclusion that \eqref{1.4} (without the introduction of general relativity) only allows for the existence of point charges. Therefore, in this section, we will first consider the electric black hole solutions in the context of electrostatic fields (hence $s=\frac{1}{2}(E^r)^2$).

\subsection{Finite energy in the Maxwell limit}\label{s3.1}

We have proved that the electric energy remains finite for any interpolation parameters $\kappa_1,\kappa_2$\cite{LY1}. In particular, when $\kappa_1\to1$ (hence $\kappa_2\to0$), the energy may still converge, which is not consistent with the ultraviolet divergence in Maxwell theory. In this subsection, we investigate whether the electric energy remains convergent in the Maxwell limit $\kappa_1\to1$, as it does in the interpolated theory.

\medskip

\textbf{Determination of electric energy for different $\kappa_1,\kappa_2$}

\begin{figure}[htbp]
    \centering
    \includegraphics[width=0.5\linewidth]{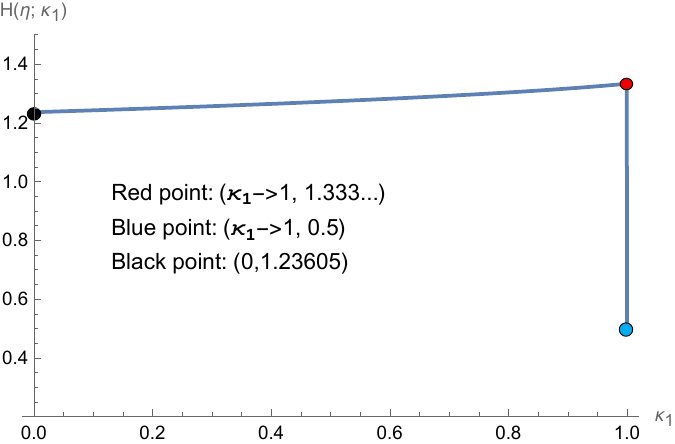}
    \caption{A plot exhibiting the normalization energy function $H(\eta;\kappa_1)$ with $\kappa_1\in[0,1)$. The black point represents the electric energy of the classical Born--Infeld theory ($\kappa_1=0$).}
    \label{fig:heta}
\end{figure}

The constitutive equation \eqref{2.32} together with the Maxwell action density \eqref{1.2} yields
\begin{align}
    \frac{q}{r^2}=\left(\kappa_1+\frac{\kappa_2}{\sqrt{1-\beta (E^r)^2}}\right)E^r,\label{3.2}
\end{align}
which admits the following asymptotic expansions of $E^r$:
\begin{align}
E^r&=\frac{1}{\sqrt{\beta}}-\frac{\kappa_2^2r^4}{2\beta^{\frac{3}{2}}q^2}-\frac{\kappa_1\kappa_2^2r^6}{\beta^2 q^3}+\frac{3(\kappa_2^2-4\kappa_1^2)\kappa_2^2r^8}{8\beta^{\frac{5}{2}}q^4}+{\rm{O}}(\kappa_2^2r^{10}),\ \ r\to0;\label{3.3}\\
E^r&=\frac{q}{r^2}-\frac{\kappa_2\beta q^3}{2r^6}-\frac{3\kappa_1\kappa_2^2\beta^2q^5}{4r^{10}}+{\rm{O}}\left(\kappa_2^2r^{-14}\right),\ \ r\to\infty.\label{3.4}
\end{align}
With $\eta=\beta(E^r)^2$, we rewrite \eqref{3.2} as
\begin{align}
    \frac{\beta q^2}{r^4}=\left(\kappa_1+\frac{\kappa_2}{\sqrt{1-\eta}}\right)^2\eta,\label{3.8}
\end{align}
and substituting this into \eqref{2.38} with $g=0$ yields
\begin{align}
    \mathcal{H}^q_{\rm MBI}=\frac1\beta\left(\frac{\kappa_1\eta}{2}+\frac{\kappa_2\eta}{\sqrt{1-\eta}(1+\sqrt{1-\eta})}\right).\label{HMBI}
\end{align}
The asymptotic expressions of $\mathcal{H}_{\rm MBI}$ obtained from \eqref{3.3} and \eqref{3.4} are
\begin{align}
    \mathcal{H}_{\rm MBI}^q&=\frac{q}{\sqrt{\beta}r^{2}}-\frac{\kappa_2}{\beta}-\frac{\kappa_1}{2 \beta}-\frac{\kappa_2^{2} r^{2}}{2 \beta^{\frac{3}{2}} q}+{\rm{O}}(r^6),\ \ r\to0,\label{mathcal{H}0}\\
    \mathcal{H}_{\rm MBI}^q&=\frac{q^{2}}{2 r^{4}}-\frac{\kappa_2q^{4}\beta}{8 r^{8}}+\frac{\kappa_2^2 \beta^{2} q^{6}}{8 r^{12}}+{\rm{O}}(r^{-16}),\ \ r\to\infty.\label{mathcal{H}infty}
\end{align}
Hence, the total electric energy (also called the energy-radius relation) is finite, given by
\begin{gather}
    \frac{E^q_{\rm MBI}}{4\pi}=\int_0^\infty\mathcal{H}(\rho)\rho^2{\rm{d}}\rho=\frac{q^2}{a}\int^1_0h(\eta,\kappa_1){\rm{d}}\eta:=\frac{q^2}{a}\cdot H(\eta;\kappa_1),\quad a=\beta^\frac14q^\frac12,\label{3.12}\\
h(\eta,\kappa_1)=\frac14\left(\frac{\kappa_1}{2}+\frac{\kappa_2}{\sqrt{1-\eta}(1+\sqrt{1-\eta})}\right)\left(\kappa_1+\frac{\kappa_2}{\sqrt{1-\eta}}\right)^{-\frac{5}{2}}\eta^{-\frac34}\left(\kappa_1+\frac{\kappa_2}{(1-\eta)^\frac32}\right),\label{3.11}
\end{gather}
where $a$ is the effective electron radius, $H(\eta;\kappa_1)$ is the normalization energy function and we have used the following quantities derived from \eqref{3.8}:
\begin{gather}
    \rho\to0,\,\,\eta\to1;\,\,\rho\to\infty,\,\,\eta\to0,\\
    \frac{{\rm d}\eta}{{\rm d}r}=-4 (\beta q^2)^{-\frac14}\left(\kappa_1+\frac{\kappa_2}{\sqrt{1-\eta}}\right)^{\frac32}\eta^{\frac54}\cdot\left({\kappa_2(1-\eta)^{-\frac32}+\kappa_1}\right)^{-1}<0.\label{3.13}
\end{gather}
It is clear that $E^q_{\rm MBI}$ depends on $\kappa_1$ and $\kappa_2$. We focus in particular on the limiting case $\kappa_1\to1$ (hence $\kappa_2\to0$), as illustrated in Fig. \ref{fig:heta}.

\medskip

\textbf{Upper bound of $H(\eta;\kappa_1)$}

Figure \ref{fig:heta} reveals a puzzling phenomenon as $\kappa\to1$: it seems that integral $H(\eta;\kappa_1)$ in \eqref{3.12} has an upper bound (red point), but its value drops sharply when $\kappa_1$ approaches $1$ extremely (blue point). In order to explain this strange phenomenon, we first note a basic property of the energy integral $H(\eta;\kappa_1)$. For any fixed $\kappa_1\in[0,1)$, $H(\eta;\kappa_1)$ is taken over a positive interval, and the integrand \eqref{3.11} is non-negative. Hence, for each fixed $\kappa_1$, the energy integral increases monotonically with the upper limit of integration; equivalently, the integral over $[0,1]$ cannot decrease as more of the positive integrand is included. Yet the numerical curve in Figure \ref{fig:heta} shows a sudden drop to $0.5$ as $\kappa_1\to1$. This drop is inconsistent with the positivity and monotonicity of $H(\eta;\kappa_1)$ at fixed $\kappa_1$, so it cannot reflect the true physical behaviour. The apparent contradiction comes from the fact that the integration region contains $\eta=1$, where the integrand becomes singular. More precisely, as $\eta\to1^-$,
\begin{align}
h(\eta;\kappa_1)\sim \kappa_2^{-\frac12}(1-\eta)^{-\frac34},
\quad \kappa_2=1-\kappa_1,\label{3.14b}
\end{align}
so when $\kappa_1\to1$ and $\kappa_2\to0$, \eqref{3.14b} diverges and the integrand blows up near $\eta=1$. This creates a \textit{boundary layer} near $\eta=1$ that is infinitely high but infinitely narrow. A naive numerical integration may miss its contribution, and the apparent value $1/2$ is then merely an artefact of the numerical procedure rather than the true physical energy. To illustrate this flaw, we present the following table:
\begin{table}[htbp]
\centering
\caption{Numerical integration of \eqref{3.12} near the boundary layer for different values of $\kappa_1$.}
\renewcommand{\arraystretch}{1.5}
\begin{tabular}{c|c|c|c|c}
\hline
$\kappa_1=$            &$1-10^{-1}$ & $1-10^{-3}$ & $1-10^{-6}$ & $1-10^{-\sigma},\sigma\ge7(\text{say})$\\
\hline
Numerical values of $H(\eta;\kappa_1)$ &1.3165809777 &1.3331274117697&1.333333125984&0.5\\
\hline
\end{tabular}
\label{tab:boundarylayer}
\end{table}

In view of Table \ref{tab:boundarylayer}, when $\kappa_2$ is less than a certain value, we cannot obtain the correct integral upper limit from $H(\eta;\kappa_1)$. To overcome this difficulty, one may either improve the accuracy of the numerical integration and carefully analyze the contribution from the boundary layer, or seek an alternative method to analyze the energy integral $H(\eta;\kappa_1)$ analytically.  Therefore, in the following content, we use an analytical approach to determine the true upper bound of the normalization energy function and verify this upper bound through more precise numerical integration.

We split $H(\eta;\kappa_1)$ into two parts, the regular integral $H_1$ and boundary-layer one $H_2$:
\begin{align}
    H(\eta;\kappa_1)=H_1+H_2,\quad H_1=\int^{1-\kappa_2^2\Lambda}_0h(\eta;\kappa_1)\,{\rm d}\eta,\quad H_2=\int_{1-\kappa_2^2\Lambda}^1h(\eta;\kappa_1)\,{\rm d}\eta.\label{3.15}
\end{align}
where $\Lambda>0$ is a constant and called the cutoff parameter. Thus, when $\eta\to1$ and $\kappa_2\to0$, the integral $H(\eta;\kappa_1)$ at the boundary layer is described by $H_2$, while $H_1$ describes the integral in the region without the boundary layer. Subsequently, we substitute $1-\eta=\kappa_2^2\xi$ into $H_2$ and obtain the integral for expanding the boundary layer integration region:
\begin{align}
    H_2=\int_0^{\Lambda}N(\xi,\kappa_1)\,{\rm d}\xi=\int_0^{\Lambda}
\frac{\left(\dfrac{\kappa_1}{2}+\dfrac{1}{\xi^{\frac12}(1+\kappa_2\xi^{\frac12})}\right)
\left(1+\kappa_1\kappa_2^2\xi^{\frac32}\right)}
{4\xi^{\frac14}\left(\kappa_1\xi^{\frac12}+1\right)^{\frac52}(1-\kappa_2^2\xi)^{\frac34}}\,{\rm d}\xi.\label{BL}
\end{align}
In view of \eqref{BL}, the cutoff parameter $\Lambda$ must be taken sufficiently large, so that the integration region of the boundary-layer integral covers the entire boundary layer and the analysis remains reliable. Moreover, at fixed $\xi$ and as $\kappa_2\to0$, the integrand of \eqref{BL} admits the pointwise limit
\begin{align}
    \lim_{\kappa_2\to0}N(\xi,\kappa_1)=\frac{2+\xi^\frac12}{8\xi^\frac34(1+\xi^\frac12)^\frac52}.
    \label{pointwiseN}
\end{align}

For each fixed $\Lambda>0$, we restrict to $0<\kappa_2^2<1/(2\Lambda)$, so that $1-\kappa_2^2\xi\ge1-\kappa_2^2\Lambda>1/2$ on $0<\xi<\Lambda$. On this fixed interval, the explicit form \eqref{BL} shows that
\begin{align}
    \bigl|N(\xi,\kappa_1)\bigr|\le C_\Lambda\,\xi^{-\frac34},
    \quad 0<\xi<\Lambda,\quad 0<\kappa_2^2<1/(2\Lambda),
    \label{eq:fixedLambdaBoundMBI}
\end{align}
with $C_\Lambda>0$ depending on $\Lambda$ but not on $\kappa_2$. Indeed, as $\xi\to0^+$ the factor $\xi^{-1/4}\le\xi^{-3/4}$ for $\xi<1$, while the remaining factors in \eqref{BL} are bounded uniformly in $\kappa_2^2\in(0,1/(2\Lambda))$ and $\xi\in(0,\Lambda)$; as $\xi\to\Lambda^-$, the factor $(1-\kappa_2^2\xi)^{-3/4}$ is bounded by $(1-\kappa_2^2\Lambda)^{-3/4}<2^{3/4}$, which is independent of $\kappa_2$ on $0<\kappa_2^2<1/(2\Lambda)$. The bound \eqref{eq:fixedLambdaBoundMBI} is integrable on $(0,\Lambda)$ and independent of $\kappa_2$.

Therefore, with $\Lambda$ held fixed and $\kappa_2\to0^+$, dominated convergence on the fixed interval $(0,\Lambda)$, with the dominating function $C_\Lambda\xi^{-3/4}$ of \eqref{eq:fixedLambdaBoundMBI}, gives
\begin{align}
    \lim_{\kappa_2\to0}H_2=\int^\Lambda_0\frac{2+\xi^\frac12}{8\xi^\frac34(1+\xi^\frac12)^\frac52}\,{\rm d}\xi+R_{\rm BL}(\Lambda),
    \label{eq:H2fixed}
\end{align}
where
\begin{align}
    R_{\rm BL}(\Lambda)=-\int_\Lambda^\infty\frac{2+\xi^\frac12}{8\xi^\frac34(1+\xi^\frac12)^\frac52}\,{\rm d}\xi=-R(\Lambda)
    \label{eq:RBLMBI}
\end{align}
is the remainder from the outer boundary-layer tail $\xi\in[\Lambda,\infty)$; it is negative and cancels $R(\Lambda)$ in \eqref{eq:Hcombined}. The integral $\int_0^\Lambda$ can be expressed as an elementary function by Chebyshev's theorem\cite{cheby1,cheby2} for any $\Lambda$.

To resolve the contribution of the boundary layer integral, we note that the full boundary layer corresponds to $\eta\in[0,1]$, i.e.\ $\xi\in[0,1/\kappa_2^2]$. Since $1/\kappa_2^2\to\infty$ as $\kappa_2\to0$, we then let
$\Lambda\to\infty$ to recover the full boundary-layer contribution:
\begin{align}
    \lim_{\Lambda\to\infty}\lim_{\kappa_2\to0}H_2
    =\int_0^\infty\frac{2+\xi^\frac12}{8\xi^\frac34(1+\xi^\frac12)^\frac52}\,{\rm d}\xi
    =\frac14\left[2B\!\left(\frac12,2\right)+B\!\left(\frac32,1\right)\right]
    =\frac56,\label{boundary layer}
\end{align}
where $B(\cdot,\cdot)$ is the Euler beta function.

\begin{remark}
Note that, since $\kappa_2^2\Lambda\in[0,1]$, one cannot take $\Lambda\to\infty$ before $\kappa_2\to0$. If $\Lambda\to\infty$ is taken first with $\kappa_2>0$ fixed, the factor $(1-\kappa_2^2\xi)^{-\frac34}$ in \eqref{BL} becomes singular at $\xi=1/\kappa_2^2$, and $H_2$ diverges at its upper end. Thus $\Lambda$ cannot be sent to infinity independently of $\kappa_2$; for every fixed $\kappa_2>0$ it is bounded by $\Lambda\le1/\kappa_2^2$. The two limits must therefore be taken in the order
\begin{align}
        \lim_{\Lambda\to\infty}\lim_{\kappa_2\to0}H_2,
\end{align}
that is, $\Lambda$ is first held fixed and $\kappa_2\to0$, and only afterwards is $\Lambda\to\infty$. In this order the bound \eqref{eq:fixedLambdaBoundMBI} is integrable on the fixed interval $(0,\Lambda)$ and independent of $\kappa_2$, so dominated convergence applies to the inner limit, and the outer limit $\Lambda\to\infty$ then recovers the full boundary-layer contribution.
\end{remark}

For the regular part $H_1$, the boundary layer has been removed by the cutoff $1-\kappa_2^2\Lambda$, so that $\eta\le 1-\kappa_2^2\Lambda<1$ strictly throughout the integration
region. Expanding $h(\eta;\kappa_1)$ near $\eta=0$ and retaining terms up to second order, we obtain
\begin{align}
    h(\eta;\kappa_1)=\eta^{-\frac34}\left[\frac18+\frac{\kappa_2}{8}\eta+\left(\frac{25\kappa_2}{128}-\frac{19\kappa_2^2}{256}\right)\eta^2+{\rm O}(\eta^3)\right],
\end{align}
where the remainder is uniform in $\kappa_2\le1$. Since $\kappa_2^2\Lambda\to0$ as $\kappa_2\to0$, the upper limit $1-\kappa_2^2\Lambda\to1$, and the integrand is regular at $\eta\to1$.
At fixed $\Lambda$, however, the interval $[1-\kappa_2^2\Lambda,1]$ still carries the outer tail of the boundary
layer and does \emph{not} shrink uniformly in $\kappa_2$; this contribution is absorbed into the remainder $R(\Lambda)$. Accordingly,
\begin{align}
    \lim_{\kappa_2\to0}H_1
    =\int_0^1\frac{{\rm d}\eta}{8\eta^\frac34}+R(\Lambda)
    =\frac12+R(\Lambda),
    \label{regular}
\end{align}
where $R(\Lambda)$ is the contribution of the outer boundary-layer tail. To derive its asymptotic behavior, we work on the fixed interval $(\Lambda,\infty)$ in the variable $\xi=(1-\eta)/\kappa_2^2$. On this interval, the same argument as for \eqref{eq:fixedLambdaBoundMBI} gives, for $0<\kappa_2^2<1/(2\Lambda)$ and $\xi>\Lambda$,
\begin{align}
    \bigl|N(\xi,\kappa_1)\bigr|\le C'_\Lambda\,\xi^{-\frac32},
    \quad \xi>\Lambda,
    \label{eq:fixedLambdaBoundMBI-tail}
\end{align}
with $C'_\Lambda>0$ independent of $\kappa_2$; indeed, for $\xi>\Lambda$ and $\kappa_2^2\xi\le\kappa_2^2\Lambda<1/2$ the factor $(1-\kappa_2^2\xi)^{-3/4}$ is bounded by $2^{3/4}$, and the large-$\xi$ behavior of \eqref{BL} is $N\sim\xi^{-3/2}$ up to a $\kappa_2$-independent constant, which is integrable on $(\Lambda,\infty)$. To apply dominated convergence on the unbounded interval $(\Lambda,\infty)$, we fix $\mathcal{K}>\Lambda$ and first consider the fixed interval $(\Lambda,\mathcal{K})$; there the integrand converges pointwise to \eqref{pointwiseN} and is dominated by $C'_\Lambda\xi^{-3/2}$, so that
\begin{align}
    \lim_{\kappa_2\to0}\int_\Lambda^\mathcal{K} N(\xi,\kappa_1)\,d\xi
    =\int_\Lambda^K\frac{2+\xi^\frac12}{8\xi^\frac34(1+\xi^\frac12)^\frac52}\,d\xi .
    \label{eq:HtailK}
\end{align}
The contribution from $\xi\in[\mathcal{K},\infty)$ is bounded by $\int_\mathcal{K}^\infty\xi^{-3/2}\,d\xi=O(\mathcal{K}^{-1/2})$, uniformly in $\kappa_2$, by \eqref{eq:fixedLambdaBoundMBI-tail}. Letting $\mathcal{K}\to\infty$ then gives
\begin{align}
    R(\Lambda)=\int_\Lambda^\infty\frac{2+\xi^\frac12}{8\xi^\frac34(1+\xi^\frac12)^\frac52}\,{\rm d}\xi={\rm O}(\Lambda^{-1}),\quad \Lambda\to\infty.
    \label{eq:RemainderMBI}
\end{align}
Table \ref{tab:Lambda} provides numerical support for the decay of $R(\Lambda)$ as $\Lambda$ increases; it does not replace the analytical estimate \eqref{eq:RemainderMBI}. Correspondingly, $\lim_{\kappa_2\to0}H_2=\frac56-R(\Lambda)$ by \eqref{eq:H2fixed} and \eqref{eq:RBLMBI}.

Combining \eqref{regular} with \eqref{boundary layer}, we obtain
\begin{align}
    \lim_{\Lambda\to\infty}\lim_{\kappa_2\to0}H(\eta;\kappa_1)
    &=\lim_{\Lambda\to\infty}\Bigl[\bigl(\tfrac12+R(\Lambda)\bigr)+\bigl(\tfrac56-R(\Lambda)\bigr)\Bigr]\notag\\
    &=\frac12+\frac56=\frac43.
    \label{eq:Hcombined}
\end{align}
The remainder cancels identically at each fixed $\Lambda$, and the limit $\Lambda\to\infty$ is taken only after the $\kappa_2\to0$ limit at fixed $\Lambda$. Consequently, as the interpolated Maxwell--Born--Infeld theory approaches pure Maxwell theory, the normalization energy remains finite and the energy-radius relation \eqref{3.12} reads
\begin{align}
    \frac{E}{4\pi}=\frac{4q^2}{3a}.\label{3.25b}
\end{align}

Now we give some numerical values to illustrate the above analytical conclusions about the boundary layer integral. First, we discuss the influence of the cutoff parameter $\Lambda$ on the integral $H(\eta;\kappa_1)$, which must be taken large enough to cover the boundary-layer region. We set $\Lambda=10^2,10^3,10^6,10^{9}$ and $\kappa_2=10^{-7}$ to summarize the values of \eqref{BL} in the following table.
\begin{table}[htbp]
    \centering
    \begin{tabular}{c|c|c|c|c}
    \hline
        $\Lambda$ & $10^2$ & $10^3$ & $10^6$ & $10^9$ \\
        \hline
        $H_1$ & 0.52433474 & 0.50784165 & 0.50024991 & 0.50000663 \\
        \hline
        $H_2$ & 0.80899857 & 0.82549166 & 0.83308341 & 0.83326686 \\
        \hline
    \end{tabular}
    \caption{For $\kappa_2=10^{-7}$, the dependence of the regular part $H_1$ and the boundary-layer part $H_2$ on the cutoff parameter $\Lambda$. As $\Lambda$ increases, $H_1$ decreases towards $1/2$ and $H_2$ increases towards $5/6$. At each finite $\Lambda$, the deviations $H_1-1/2$ and $5/6-H_2$ reflect, numerically, the equal and opposite remainder $R(\Lambda)$ in \eqref{eq:RemainderMBI}.}
    \label{tab:Lambda}
\end{table}

With the analytical analysis and the above numerical illustration of the effect of $\Lambda$ at hand, we can now extract the upper bound of the electric energy as $\Lambda$ from the numerical side. The results are presented in Table \ref{tab:energylimit}, which extends Table \ref{tab:boundarylayer}.
\begin{table}[htbp]
\centering
\renewcommand{\arraystretch}{1.5}
\begin{tabular}{c|c|c|c|c|c}
\hline
$\kappa_1, \Lambda$&1--$10^{-9},10^2$ &1--$10^{-12},10^5$ & 1--$10^{-15},10^8$ & 1--$10^{-18},10^{11}$&1--$10^{-21},10^{14}$\\
\hline
Numerical values of $H_2$ &0.80899856&0.83254339&0.83330833&0.83333254&0.83333331\\
\hline
\end{tabular}
\caption{Numerical integration of \eqref{3.12} near the boundary layer for different values of $\kappa_1$. It should be noted that, as $\kappa_2$ decreases, the cutoff parameter $\Lambda$ must be increased accordingly so as to capture as much of the boundary-layer contribution as possible. At each finite $\Lambda$, the tabulated $H_2$ differs from $5/6$ by the remainder $R(\Lambda)$ in \eqref{eq:RemainderMBI}.}
\label{tab:energylimit}
\end{table}

\subsection{Interpolated electric black hole solution}

We now construct the interpolated black hole solution via Taylor expansions and the implicit function $\eta$. Substituting \eqref{HMBI}--\eqref{mathcal{H}infty} into \eqref{2.45} and \eqref{2.47} yields
\begin{align}
    A_{\rm MBI}^q(r)&=\begin{cases}
        1-\dfrac{2G(M-E^q_{\rm MBI})}{r}-\dfrac{8\pi Gq}{\sqrt{\beta}}+\dfrac{4\pi G(\kappa_1+2\kappa_2)r^2}{3\beta}+{\rm{O}}(r^4),\\
        1-\dfrac{2G(M-E^q_{\rm MBI})}{r}-\dfrac{8\pi G}{\beta r}\displaystyle\int^r_0\left(\dfrac{\kappa_1\eta(\rho)}{2}+\dfrac{\kappa_2\eta(\rho)}{\sqrt{1-\eta(\rho)}(1+\sqrt{1-\eta(\rho)})}\right)\rho^2{\rm d}\rho,
    \end{cases}\ \ r\to0,\label{3.21}\\
    A_{\rm{MBI}}^q(r)&=\begin{cases}
        1-\dfrac{2GM}{r}+\dfrac{4 \pi Gq^{2}}{r^{2}}-\dfrac{\kappa_2\pi G\beta q^{4}}{5 r^{6}}+{\rm{O}}(r^{-10}),\\
        1-\dfrac{2GM}{r}+\dfrac{8\pi G}{r}\displaystyle\int^\infty_r\left(\dfrac{\kappa_1\eta(\rho)}{2}+\dfrac{\kappa_2\eta(\rho)}{\sqrt{1-\eta(\rho)}(1+\sqrt{1-\eta(\rho)})}\right)\rho^2{\rm d}\rho,
    \end{cases}\ \ r\to\infty.\label{3.22}
\end{align}
From \eqref{3.8}, we have
\begin{align}
    \rho\to0,\quad\eta\to1;\quad\rho\to r,\quad\eta\to\eta_0;\quad\rho\to\infty,\quad \eta\to0.\label{3.14}
\end{align}
Here, $\eta_0$ denotes the value of $\eta$ at a distance $\rho=r$. Using \eqref{3.14}, we can rewrite the second lines of \eqref{3.21} and \eqref{3.22} as
\begin{align}
    A_{\rm MBI}^q(r)&=1-\dfrac{2G(M-E^q_{\rm MBI})}{r}-\dfrac{2\pi Gq^\frac32}{\beta^\frac14r}\int^1_{\eta_0} h(\eta,\kappa_1){\rm d}\eta,\quad r\ll1,\\
    A_{\rm{MBI}}^q(r)&=1-\dfrac{2GM}{r}+\dfrac{2\pi Gq^\frac32}{\beta^\frac14r}\int^{\eta_0}_0 h(\eta,\kappa_1){\rm{d}}\eta,\quad r\gg1.
\end{align}

\medskip

\textbf{Kretschmann curvature}

The curvature singularity at $r=0$ can be analyzed via the Kretschmann curvature
\begin{align}
    K=R_{\mu\nu\alpha\beta}R^{\mu\nu\alpha\beta}=\frac{(r^2{\cal A}^{\prime\prime})^2+4(r{\cal A}^\prime)^2+4({\cal A}-1)^2}{r^4},\ \ {\cal A}=A_{\rm MBI}^q(r),\ \ \prime=\frac{{\rm{d}}}{{\rm{d}}r},\label{formula of Kretschmann}
\end{align}
which, using \eqref{3.21}, evaluates to
\begin{align}
    K=&\frac{48G^2(M-E^q_{\rm MBI})^2}{r^6}+\frac{128\pi G^2q(M-E^q_{\rm MBI})}{\sqrt{\beta}r^5}+\frac{256\pi^2G^2q^2}{\beta r^4}+\frac{64\pi G^2(M-E^q_{\rm MBI})(\kappa_1+2\kappa_2)}{3\beta r^4}+{\rm{O}}(r^{-2}).\label{Kretschmann of electric}
\end{align}
Thus, $K$ exhibits a Schwarzschild-like $r^{-6}$ divergence. However, under the critical mass-energy condition
$M=E_{\rm MBI}^q$ (i.e., when the gravitational mass equals the energy) \cite{brewin2007simple}, the leading $r^{-6}$ term vanishes and $K\sim r^{-4}$, so the singularity is relegated. This shows that the singularity structure is robust against the specific values of $\kappa_1$ and $\kappa_2$.

\medskip

\textbf{Interpolated magnetic and dyonic black hole solutions}

Although the energies of magnetic monopoles and dyons always exhibit explosive behavior near the origin in our interpolated theory \cite{LY1}, we can still derive the energy behavior as $r\to\infty$. This exploration aims to make a comparison with the classical Maxwell theory: despite the divergent energy of point charges in Maxwell theory as $r\to0$, explicit black hole solutions still exist in the infinite region.

In the setting of a pure magnetic field with ${E^r}=0$ and $s=-{g^2}/{(2r^4)}$, the Hamiltonian energy density reads
\begin{align}
    \mathcal{H}_{\rm MBI}^g=&\frac{\kappa_1g^2}{2r^4}-\frac{\kappa_2}{\beta}\left(1-\sqrt{1+\frac{\beta g^2}{r^4}}\right)\notag\\
    =&\frac{g^2}{2r^4}-\frac{\kappa_2\beta g^4}{8r^8}+\frac{\kappa_2\beta^2 g^6}{16r^{12}}-\frac{5\kappa_2\beta^3 g^8}{128r^{16}}+{\rm O}(r^{-20}),
\end{align}
such that the magnetic energy near infinity is given by
\begin{align}
    \frac{E_{\rm MBI}^g}{4\pi}=\int^\infty_r\mathcal{H}_{\rm MBI}^g(\rho)\rho^2{\rm d}\rho=\begin{cases}
        \dfrac{g^2}{2r}-\dfrac{\kappa_2\beta g^4}{40r^5}+\dfrac{\kappa_2\beta^2g^6}{144r^9}+{\rm O}(r^{-13}),\\
        \dfrac{\kappa_1g^2}{2r}+\dfrac{\kappa_2r^3}{3\beta}-\dfrac{\kappa_2r^3}{3\beta}\cdot_{2}{F}_1\left[-\dfrac34,-\dfrac12,\dfrac14,-\dfrac{\beta g^2}{r^4}\right],
    \end{cases}\quad r\gg1,
\end{align}
leading to the magnetic black hole solution obtained from \eqref{2.47}:
\begin{align}
    A^{g}_{\rm MBI}(r)=
    \begin{cases}
        1-\dfrac{2GM}{r}+{4\pi Gg^2}\left(\dfrac{1}{r^2}-\dfrac{\kappa_2\beta g^2}{20r^6}+\dfrac{\kappa_2\beta^2g^4}{72r^{10}}-\dfrac{5\kappa_2\beta^3g^6}{832r^{14}}\right)+{\rm O}(r^{-18}),\\
        1-\dfrac{2GM}{r}+\dfrac{4\pi Gg^2\kappa_1}{r^2}+\dfrac{8\pi G\kappa_2r^2}{3\beta}-\dfrac{8\pi G\kappa_2r^2}{3\beta}\cdot_{2}{F}_1\left[-\dfrac34,-\dfrac12,\dfrac14,-\dfrac{\beta g^2}{r^4}\right],
    \end{cases}\quad r\gg1,\label{3.31}
\end{align}
where $\prescript{ }{2}{F}_1[a_1,a_2,a_3,z]$ is the Gauss hypergeometric function, which takes complex values when $|z|\ge1$.

For a dyonic configuration with both electric charge $q>0$ and magnetic charge $g>0$, we solve \eqref{2.32} for $r\to\infty$ to obtain
\begin{align}
    E^r=\frac{q}{r^2}-\frac{\kappa_2\beta q^3}{2r^6}\left(1-\frac{g^2}{q^2}+2\kappa_2\frac{g^2}{q^2}\right)+\frac{\kappa_2\beta^2 q^5}{8r^{10}}\left(4+\kappa_1-\kappa_2\right)+{\rm O}(r^{-14}).
\end{align}
Substituting into \eqref{2.38} yields the asymptotic dyonic Hamiltonian density
\begin{align}
    \mathcal H_{\rm MBI}^{q,g}=\frac{q^2+g^2}{2r^4}-\frac{\kappa_2\beta(q^4+g^4)}{8r^8}-\frac{\kappa^2(\kappa_1+\frac12)q^2g^2}{2r^8}+{\rm O}(r^{-12}),\quad r\to\infty.
\end{align}
Consequently, from \eqref{2.31}, the dyonic black hole metric function is
\begin{align}
    A_{\rm MBI}^{q,g}(r)&=1-\frac{2GM}{r}+\frac{4\pi G(q^2+g^2)}{r^2}-\frac{\pi G\left[\beta\kappa_2(q^4+g^4)+4\kappa^2(\kappa_1+\frac12)q^2g^2\right]}{5r^6}+{\rm O}(r^{-10}),\quad r\to\infty.
    \label{3.32}
\end{align}

\textbf{Comparison among black hole solutions}

In the limiting case $\kappa_2=0$ and $\kappa_1=1$, the model \eqref{1.4} reduces to the pure Maxwell theory. Our interpolated theory approximates the pure Maxwell theory. On the one hand, \eqref{3.3} yields
\begin{align}
    E^r\to\frac{1}{\sqrt{\beta}},\quad r\to0,\quad\kappa_1\to1,\label{3.29}
\end{align}
which reveals a signature of $\kappa_2\in(0,1]$: even when the classical Born--Infeld theory carries a negligible weight in the interpolation, it can still play a crucial role in preventing the electric field from diverging near the origin. Moreover, \eqref{3.29} admits a clear interpretation in terms of the free charge distribution: as $\kappa_1\to1$ ($\kappa_2\to0$), the Maxwell term becomes dominant, and the ratio $q_{\rm free}/q$ approaches unity for $R\ge a$; that is, almost all free charge is captured within the effective radius $a$. Consequently, the electric field near the origin saturates at the Born--Infeld maximal value $1/\sqrt{\beta}$, even though the Born--Infeld weight $\kappa_2$ is arbitrarily small. This shows that the distribution of free charge, the saturation of the electric field, and the dominance of the Maxwell term are mutually consistent features of the interpolated model.

On the other hand, for large $r$, the electric field behaves as $E^r=\frac q{r^2}$, and \eqref{3.22} recovers the Reissner--Nordström solution
\begin{align}
    A_{\rm{RN}}(r)=1-\frac{2GM}{r}+\frac{4\pi Gq^2}{r^2},\label{A(r)RN}
\end{align}
since $\mathcal{H}_{\rm Maxwell}=\frac12(E^r)^2=\frac\eta{2\beta}$ in our notation. By contrast, the classical Born--Infeld black hole solution takes the form \cite{Y-book,Hoffmann}
\begin{align}
A^q_{\rm{BI}}(r)&=\begin{cases}
    1-\dfrac{2GM}{r}+\dfrac{4\pi Gq^{2}}{r^{2}}-\dfrac{\pi G\beta q^{4}}{5 r^{6}}+{\rm{O}}(r^{-10}),\\
    1-\dfrac{2GM}{r}+\dfrac{8\pi Gq^2}{r}\displaystyle\int^\infty_r\frac{{\rm d}\rho}{\rho^2+\sqrt{\rho^4+\beta q^2}},
\end{cases}\quad r\to\infty;\label{A(r)BI}\\
A^q_{\rm{BI}}(r)&=1-\dfrac{2G(M-E_{\rm BI})}{r}-\dfrac{8\pi G}{r}{\displaystyle\int^r_0}\dfrac{1}{\beta}\left(\dfrac{\eta(\rho)}{\sqrt{1-\eta(\rho)}(1+\sqrt{1-\eta(\rho)})}\right)\rho^2{\rm d}\rho,\quad r\to0.\label{A(r)BI2}
\end{align}
with the associated Born--Infeld Hamiltonian energy density
\begin{align}
    \mathcal{H}_{\rm BI}=\frac1\beta\left((1-\eta)^{-\frac12}-1\right).\label{HBI}
\end{align}
In view of the asymptotic expressions in \eqref{3.22} and \eqref{A(r)BI}, we observe that the interpolated solution lies between the Born--Infeld and Reissner--Nordström cases (see \eqref{A(r)RN}):
\begin{align}
    A_{\rm{BI}}^q(r)\le A^q_{\rm{MBI}}(r)\le A_{\rm{RN}}(r),\quad r\to\infty.\label{3.25}
\end{align}

Indeed, we note that \eqref{HMBI} satisfies
\begin{gather}
\frac{\partial\mathcal{H}_{\rm MBI}(\kappa_1)}{\partial\kappa_1}=\frac\eta\beta\left(\frac12-\frac{1}{\sqrt{1-\eta}(1+\sqrt{1-\eta})}\right)\le0,\quad \eta\in[0,1),
\end{gather}
which, together with \eqref{HBI} and $\mathcal{H}_{\rm Maxwell}=\frac\eta{2\beta}$, implies that the Hamiltonian energy density of \eqref{1.4} is also interpolated between $\mathcal{H}_{\rm BI}$ and $\mathcal{H}_{\rm Maxwell}$:
\begin{gather}
    \mathcal{H}_{\rm Maxwell}\le\mathcal{H}_{\rm MBI}(\kappa_1)\le\mathcal{H}_{\rm BI}\iff\frac\eta{2}\le\left(\frac{\kappa_1\eta}{2}+\frac{\kappa_2\eta}{\sqrt{1-\eta}(1+\sqrt{1-\eta})}\right)\le\frac{\eta}{\sqrt{1-\eta}(1+\sqrt{1-\eta})},\label{3.27}\\
    \mathcal{H}_{\rm Maxwell}=[\mathcal{H}_{\rm MBI}(\kappa_1)]_{\rm minimum}=\mathcal{H}_{\rm MBI}(1),\quad \mathcal{H}_{\rm BI}=[\mathcal{H}_{\rm MBI}(\kappa_1)]_{\rm maximum}=\mathcal{H}_{\rm MBI}(0).
\end{gather}
To clearly illustrate the influence of $\kappa_1$ on these Hamiltonians, we present the following plot.
\begin{figure}[htbp]
    \centering
    \includegraphics[width=0.5\linewidth]{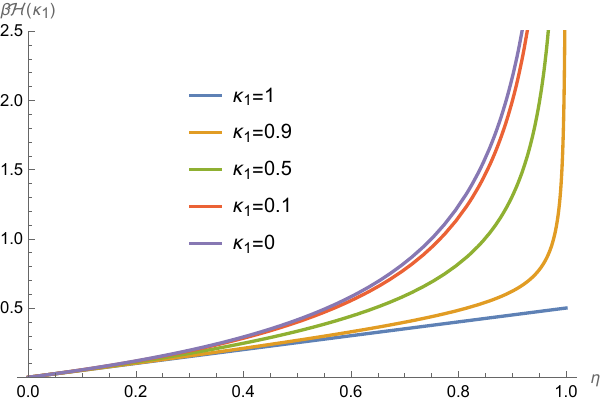}
    \caption{A plot shows the monotonicity of $\mathcal{H}$ in terms of $\kappa_1=1,0.9,0.5,0.1,0$. The Hamiltonian of the interpolation theory is interpolated in Maxwell and the classical Born--Infeld theories.}
    \label{fig:betaHkappa1}
\end{figure}

Hence, in the electrostatic field, the black hole solutions of the interpolation model smoothly interpolate between the Reissner--Nordström and classical Born--Infeld solutions.

Furthermore, in the magnetostatic field, the magnetic black hole solution \eqref{3.31} yields some observations:
\begin{enumerate}
    \item When $\kappa_1=1$ and $\kappa_2=0$, the interpolated model reduces to Maxwell formalism, and \eqref{3.31} becomes
    \begin{align}
        A^g_{\rm MBI}(r)=1-\frac{2GM}{r}+\frac{4\pi Gg^2}{r^2},
    \end{align}
    which demonstrates a symmetry between the Maxwell electric black hole solution and the magnetic black hole solution of the interpolated model.
    \item When $\kappa_1=0$ and $\kappa_2=1$, the interpolated model reduces to the classical Born--Infeld theory, as the latter possesses electromagnetic symmetry in the black hole solution. Hence, we present the following inequality between the magnetic black hole solutions:
    \begin{align}
        A^g_{\rm BI}(r)\le A^g_{\rm MBI}(r).
    \end{align}
\end{enumerate}

\subsection{Energy conditions}

The above analysis has established the structural properties of the interpolated Maxwell--Born--Infeld black holes, including the asymptotic expansions, free charge distribution, finite total energy, Kretschmann curvature, and horizon existence conditions. However, the physical viability of a solution in general relativity further requires that it respect suitable energy conditions\cite{hawking1973}, which encode basic expectations such as the positivity of energy density and the attractive nature of gravity. In the context of nonlinear electrodynamics, these conditions translate into concrete inequalities involving the Lagrangian $\mathcal{L}(s)$ and its derivatives\cite{Dymnikova1,Dymnikova2,yang2022electromagnetic}, given by
\begin{gather}
    \text{Weak energy condition(WEC): }\mathcal{L}'(s)(E^r)^2-\mathcal{L}(s)\ge0,\quad \mathcal{L}'(s)\left((E^r)^2+r^4B(r)^2\right)\ge0,\label{WEC}\\
    \text{Dominant energy condition(DEC): }\mathcal{L}'(s)\left((E^r)^2-r^4B(r)^2\right)-2\mathcal{L}(s)\ge0,\label{DEC}\\
    \text{Strong energy condition(SEC): }r^4\mathcal{L}'(s)B(r)^2+\mathcal{L}(s)\ge0,\label{SEC}
\end{gather}
where $B(r)$ is given by \eqref{2.23}. The extent to which the energy conditions hold is closely tied to spacetime regularity: violations may signal the appearance or disappearance of singularities, or point to exotic matter that could destabilize the black hole.

In what follows, we examine the weak, dominant, and strong energy conditions for the interpolated model \eqref{1.4}, and determine the constraints they impose on the parameters $\kappa_1$, $\kappa_2$ and the field strength.

First, we examine the pure electric field case with $s=\frac{1}{2}(E^r)^2$ and $B^r=0$. Actually, the first condition of \eqref{WEC} is the Hamiltonian energy density $T_{00}$, which is always non-negative in the electrostatic field. Inserting \eqref{1.4} into \eqref{WEC}-\eqref{SEC}, we find that all the energy conditions are valid:
\begin{align}
    \text{WEC: }&\mathcal{H}\ge0,\quad \left(\kappa_1+\frac{\kappa_2}{\sqrt{1-\beta(E^r)^2}}\right)(E^r)^2\ge0,\\
     \text{DEC: }&\frac{\kappa_2}{\beta}\left(\frac{\beta (E^r)^2}{\sqrt{1-\beta(E^r)^2}}+2\sqrt{1-\beta(E^r)^2}-2\right)\ge0,\\
     \text{SEC: }& \mathcal{L}_{\rm MBI}\ge0.
\end{align}

Next, we turn to the pure magnetic field case with $s=-\frac{1}{2}(B^r)^2$. Although monopoles and dyons have been excluded from the interpolated theory on energetic grounds in \cite{LY1}, it is still of considerable interest to examine the energy conditions in a magnetic background. Such an analysis provides an independent, multi-faceted perspective on why the interpolated model does not accommodate magnetic charges.

For a pure magnetic field the invariants are $E^r=0$ and
$\mathcal B^2:=r^4(B^r)^2={g^2}/{r^4}$, so that
$s=-\tfrac12\mathcal B^2=-\tfrac{\zeta}{2\beta}$ with
$\zeta(r)=\beta g^2/r^4$, and the Lagrangian and its derivative read
\begin{align}
\mathcal{L}_{\rm MBI}=-\frac{\kappa_1\zeta}{2\beta}+\frac{\kappa_2}{\beta}\left(1-\sqrt{1+\zeta}\right),\quad
\mathcal{L}'(s)=\kappa_1+\frac{\kappa_2}{\sqrt{1+\zeta}}>0.\label{magL}
\end{align}
The energy density and the tangential pressure are
\begin{align}
\rho=-\mathcal{L}_{\rm MBI}=\frac1\beta\left[\frac{\kappa_1\zeta}{2}+\kappa_2\left(\sqrt{1+\zeta}-1\right)\right]\ge0,\quad
p_\perp=\mathcal{L}_{\rm MBI}+\mathcal{L}'(s)\mathcal B^2
=\frac1\beta\left[\frac{\kappa_1\zeta}{2}+\kappa_2\left(\frac{\zeta}{\sqrt{1+\zeta}}+1-\sqrt{1+\zeta}\right)\right]\ge0,
\label{magrhopp}
\end{align}
where the non-negativity in \eqref{magrhopp} follows since $\sqrt{1+\zeta}\ge1$ and $\frac{\zeta}{\sqrt{1+\zeta}}+1-\sqrt{1+\zeta}=\frac{\sqrt{1+\zeta}-1}{\sqrt{1+\zeta}}\ge0$.

For the weak energy condition, the first inequality of \eqref{WEC} is $\rho\ge0$, manifestly true by \eqref{magrhopp}; the second is $\mathcal L'(s)\mathcal B^2\ge0$, true since $\mathcal L'(s)>0$. Hence the WEC holds everywhere in the pure magnetic case.

The dominant energy condition \eqref{DEC} reduces to $\rho-p_\perp\ge0$. Using \eqref{magL}--\eqref{magrhopp} and $\beta\mathcal B^2=\zeta$, we obtain
\begin{align}
\rho-p_\perp=-2\mathcal{L}_{\rm MBI}-\mathcal L'(s)\mathcal B^2
=\frac{\kappa_2}{\beta}\left(\sqrt{1+\zeta}-2+\frac{1}{\sqrt{1+\zeta}}\right)
=\frac{\kappa_2}{\beta}\frac{\left(\sqrt{1+\zeta}-1\right)^2}{\sqrt{1+\zeta}}\ge0,
\label{magDEC}
\end{align}
which holds identically, with equality only at $\zeta=0$ (i.e.\ $g=0$). Thus the DEC is also satisfied everywhere.

For the strong energy condition, the SEC \eqref{SEC} reads $\mathcal L'(s)\mathcal B^2+\mathcal{L}_{\rm MBI}\ge0$, i.e.\ $p_\perp\ge0$, which is manifest from \eqref{magrhopp}. Hence the SEC is satisfied everywhere as well.

We now extend the analysis to the dyonic situation where both $E^r$ and $B^r$ are non-vanishing. Setting $\eta=\beta(E^r)^2\in[0,1)$ and $\zeta=\beta g^2/r^4\ge0$, so that $w:=1-\eta+\zeta$, the action density becomes $s=(\eta-\zeta)/(2\beta)$ and the energy density and tangential pressure read
\begin{align}
\rho&=\frac1\beta\left[\frac{\kappa_1}{2}(\eta+\zeta)+\kappa_2\left(\frac{1+\zeta}{\sqrt{w}}-1\right)\right]\ge0,\\
p_\perp&=\frac1\beta\left[\frac{\kappa_1}{2}(\eta+\zeta)+\kappa_2\left(\frac{\zeta}{\sqrt{w}}+1-\sqrt{w}\right)\right]\ge0.
\end{align}
The weak and strong energy conditions are therefore satisfied for all admissible $\eta$ and $\zeta$, since 
\begin{gather}
    \frac{1+\zeta}{\sqrt w}-1=\frac{1+\zeta-\sqrt w}{\sqrt w}\ge0,\\
    \frac{\zeta}{\sqrt w}+1-\sqrt w=\frac{\zeta-\sqrt w(\sqrt w-1)}{\sqrt w}=\frac{\sqrt w-(1-\eta)}{\sqrt w}\ge0.
\end{gather}
For the dominant energy condition,
\begin{align}
\rho-p_\perp=\frac{\kappa_2}{\beta}\left(\frac{1}{\sqrt w}-2+\sqrt w\right)
=\frac{\kappa_2}{\beta}\frac{\left(1-\sqrt{1-\eta+\zeta}\right)^2}{\sqrt{1-\eta+\zeta}}\ge0,
\end{align}
with equality only at $s=0$ ($\eta=\zeta$). Hence the DEC is also satisfied, and no energy condition is violated in the dyonic sector.

In summary, the interpolated Maxwell--Born--Infeld theory satisfies all three classical energy conditions--weak, dominant, and strong--in the pure electric, pure magnetic, and dyonic sectors alike. The exclusion of magnetic monopoles and dyons is therefore not signaled by any violation of the energy conditions; rather, it is encoded entirely in the divergence of their total energy, consistent with the flat-spacetime analysis of \cite{LY1}. In this respect the energy conditions play a diagnostic rather than a veto role: their validity rules out the exotic-matter pathologies that would otherwise obscure the purely energetic obstruction against magnetic and dyonic configurations.

\medskip

\section{Event horizons of the interpolated black holes}\label{s4}

In Section \ref{s3}, we constructed a family of interpolated charged black hole solutions. It is therefore of considerable interest to analyze and compare the horizon structures of the interpolated black holes.

We begin with the Reissner--Nordstr\"{o}m black hole. Its event horizons are determined by $A_{\rm RN}(r)=0$, which yields
\begin{align}
    r=GM\pm\sqrt{G^2M^2-4\pi Gq^2}.
\end{align}
If $GM^2 < 4\pi q^2$, the equation admits no real positive root, and the spacetime exhibits a naked singularity, thereby violating the Cosmic Censorship Hypothesis. Otherwise, the horizon structure is characterized as follows:
\begin{align}
    & GM^2 = 4\pi q^2: \quad \text{a single event horizon } r_h = GM; \\
    & GM^2 > 4\pi q^2: \quad \text{outer and inner horizons,}
    \begin{cases}
        r_+=GM+\sqrt{G^2M^2-4\pi Gq^2}, \\
        r_-=GM-\sqrt{G^2M^2-4\pi Gq^2}.
    \end{cases}
\end{align}
This observation motivates a detailed investigation into the number of event horizons for the electric black hole in the interpolated theory, and an identification of the parameters that govern the transitions between the different horizon configurations. 

\subsection{Existence and multiplicity of event horizons in general interpolated theories}\label{s4.1}

We first analyze the existence and multiplicity of the Maxwell--Born--Infeld electric black holes. It is worth noting that, when nonlinear electrodynamics is coupled to general relativity, the resulting black hole solutions acquire numerous higher-order correction terms relative to the Reissner--Nordstr\"{o}m black hole, which makes an explicit analytic determination of the zeros of $A(r)$—and hence the event horizons—extremely difficult. In much of the existing literature\cite{Y-dyonic2022,Gao2021,yang2022electromagnetic,LiuZhang2025}, certain parameters are fixed or graphical methods are employed to infer the presence of horizons. Such analyses do yield reliable qualitative conclusions, but they do not fully capture the intrinsic structural features that the nonlinearity brings in. Unless the parameter restrictions are lifted, one can at best characterise the event horizon approximately, and its precise multiplicity cannot be determined. The present subsection therefore aims to analyze the existence and exact number of event horizons from the structure of black hole solutions.

\medskip

 \textbf{Monotonicity of $rA(r)$}

Since $rA^q_{\rm MBI}(r)$ enjoys the same zeros as $A^q_{\rm MBI}(r)$, it suffices to investigate the monotonicity of the former. Using \eqref{1.4} and the second line of \eqref{3.22}, we obtain
\begin{align}
    (rA^q_{\rm MBI}(r))'
    &= \frac{\rm d}{{\rm d}r}\left(r-2GM+8\pi G\int^\infty_r\mathcal{H}^q_{\rm MBI}(\rho)\rho^2\,\mathrm d\rho\right)\notag\\
    &=1-8\pi Gr^2\mathcal{H}^q_{\rm MBI}(r)\notag\\
    &=1-\frac{8\pi Gq}{\sqrt{\beta}}\cdot\mathcal{A}^q_{\rm MBI}(\eta,\kappa_1),\label{4.4}
\end{align}
where we call $\mathcal{A}^q_{\rm MBI}(\eta,\kappa_1)$ the \textit{horizon monotonicity function}, which depends on two independent variables $\eta\in[0,1)$ and $\kappa_1\in[0,1)$, given by
\begin{align}
   \mathcal{A}^q_{\rm MBI}(\eta,\kappa_1):= \frac{\kappa_1\eta\sqrt{1-\eta}+2\kappa_2(1-\sqrt{1-\eta})}
{2\sqrt{\eta}\left(\kappa_1\sqrt{1-\eta}+\kappa_2\right)},\quad\kappa_2=1-\kappa_1.
\end{align}
Since
\begin{align}
    \frac{\partial \mathcal A^q_{\rm MBI}}{\partial \kappa_1}
    &= -\frac{(1-\sqrt{1-\eta})^2\sqrt{1-\eta}}
    {2\sqrt{\eta}\left(\kappa_1\sqrt{1-\eta}+\kappa_2\right)^2}<0,\label{dAqdk1}\\[4pt]
    \frac{\partial \mathcal A^q_{\rm MBI}}{\partial \eta}
    &= \frac{\left[\kappa_1(1+\sqrt{1-\eta})+2\kappa_2\right]
    \left[\kappa_1(1-\eta)^{\frac32}+\kappa_2\right]}
    {4\eta\,(1+\sqrt{1-\eta})\left(\kappa_1\sqrt{1-\eta}+\kappa_2\right)^2}>0,\label{dAqdeta}
\end{align}
we see that $\mathcal A^q_{\rm MBI}$ is strictly decreasing in $\kappa_1$ and strictly increasing in $\eta$ on the domain $(\eta,\kappa_1)\in[0,1)\times[0,1)$. In particular it admits no interior stationary points, so its extreme values are attained only in the limits $\eta\to0^+$ and $\eta\to1^-$:
\begin{align}
    \lim_{\eta\to1^-}\mathcal A^q_{\rm MBI}(\eta,\kappa_1)
    =\frac{2\kappa_2}{2\kappa_2}=1,\quad
    \lim_{\eta\to0^+}\mathcal A^q_{\rm MBI}(\eta,\kappa_1)=0,
\end{align}
for every fixed $\kappa_2\in(0,1]$. Hence
\begin{align}
    0\le\mathcal{A}^q_{\rm MBI}(\eta,\kappa_1)<1.
\end{align}
Substituting this into \eqref{4.4}, we have
\begin{align}
    \mathcal{C}<(rA^q_{\rm MBI}(r))'\le1,\quad \mathcal{C}\equiv1-\frac{8\pi Gq}{\sqrt{\beta}}.\label{rAprime}
\end{align}

The monotonicity of $rA^q_{\rm MBI}(r)$ is therefore governed by the sign of $\mathcal{C}$, which inspires us to consider the following two cases.

\medskip

 \textbf{Case 1: $\mathcal{C}\ge0$.}

In this case, \eqref{rAprime} implies $(rA^q_{\rm MBI}(r))'>0$, so $rA^q_{\rm MBI}(r)$ is strictly increasing on $(0,\infty)$. Its infimum is attained as $r \to 0$. From \eqref{3.22} and \eqref{3.12}, we have
\begin{align}
    \inf_{r>0}rA^q_{\rm MBI}(r)
    &=\lim_{\substack{r\to0\\\eta\to1}}
       \left(r-2GM+\frac{2\pi G q^{\frac32}}{\beta^{\frac14}}
       \int^\eta_0 h(\tau,\kappa_1)\mathrm d\tau\right) \notag \\
    &=2G(E^q_{\rm MBI}-M).
\end{align}
Consequently, the existence of horizons is determined solely by the mass--energy relation:
\begin{enumerate}
    \item If $M<E^q_{\rm MBI}$, then $rA^q_{\rm MBI}(r)>0$ for all $r>0$, and hence no event horizon exists.
    \item If $M=E^q_{\rm MBI}$, then $\inf \left[rA^q_{\rm MBI}(r)\right]=0$, attained only at $r=0$, corresponding to a naked singularity.
    \item If $M>E^q_{\rm MBI}$, then there is exactly one event horizon.
\end{enumerate}

\medskip

\textbf{Case 2: $\mathcal{C}<0$.}

In this case, $rA^q_{\rm MBI}(r)$ is not monotonic. It possesses a unique minimum at some $r_1 \in (0,\infty)$, decreasing on $(0,r_1)$ and increasing on $(r_1,\infty)$. To determine the number of horizons, we must examine the sign of $rA^q_{\rm MBI}(r)$ over the entire interval, taking into account both the endpoint behaviour and the value at the minimum.

The asymptotic behaviour is as follows. As $r \to \infty$, we have $\eta\to0$ and $\mathcal{H}^q_{\rm MBI}\to0$ in \eqref{HMBI}; hence \eqref{3.22} gives $rA^q_{\rm MBI}(r)\to+\infty$. As $r\to0$, we have $\eta\to1$, yielding
\begin{align}
    rA^q_{\rm MBI}(r)\to-2G(M-E^q_{\rm MBI})
    \begin{cases}
        >0, & M<E^q_{\rm MBI},\\
        =0, & M=E^q_{\rm MBI},\\
        <0, & M>E^q_{\rm MBI},
    \end{cases}
    \quad r \to 0.\label{rAqMBI0}
\end{align}

Based on these properties, the horizon structure can be classified as follows:
\begin{table}[htbp]
\centering
\caption{Horizon classification for the interpolated black hole when $\sqrt{\beta}<{8\pi Gq}$.}
\renewcommand{\arraystretch}{1.5}
\begin{tabular}{c|c|c}
\hline
$r_1A^q_{\rm MBI}(r_1)$ & Mass-energy condition & Horizon type \\
\hline
$>0$ & $M< E^q_{\rm MBI},-2G(M-E)>r_1A(r_1)$ & 0 (naked singularity) \\
$>0$ & $M< E^q_{\rm MBI},-2G(M-E)\le r_1A(r_1)$ & unphysical \\
$>0$ & $M\ge E^q_{\rm MBI}$ & unphysical \\
\hline
$=0$ & $M<E^q_{\rm MBI}$ & 1 at $r_1$ \\
$=0$ & $M\ge E^q_{\rm MBI}$ & unphysical \\
\hline
$<0$ & $M<E^q_{\rm MBI}$ & 2 (inner and outer) \\
$<0$ & $M=E^q_{\rm MBI}$ & 1 \\
$<0$ & $M>E^q_{\rm MBI},\; r_1A^q_{\rm MBI}(r_1)<-2G(M-E)$ & 1 \\
$<0$ & $M>E^q_{\rm MBI},\; r_1A^q_{\rm MBI}(r_1)\ge-2G(M-E)$ & unphysical \\
\hline
\end{tabular}
\label{tab:MBIhorizon}
\end{table}

Comparing the above classification with the Reissner--Nordstr\"{o}m case, we observe that the interpolated Maxwell--Born--Infeld black holes exhibit a significantly richer horizon structure. The number and nature of the horizons depend on a more intricate interplay of the parameters $\beta$, $G$, $M$, and $E^q_{\rm MBI}$, in marked contrast to the relatively simple Reissner--Nordstr\"{o}m scenario. This influence of the classical Born--Infeld theory manifests itself through the parameter $\beta$, which distinguishes the monotonicity properties of $rA^q_{\rm MBI}(r)$. It is therefore of interest to examine how the interpolation parameters and the Born parameter $b=\beta^{-\frac12}$ affect the size of the event horizon. In the next subsection, we take the horizon-existence examples classified above as concrete cases: on the one hand, they provide explicit numerical illustrations of the theoretical classification; on the other hand, they show how the allocation of weights between the Maxwell and Born--Infeld terms in the interpolated theory, together with the Born parameter, affects the size of the event horizon.

\begin{remark}\label{remark:unphysical}
In Table \ref{tab:MBIhorizon} the entries labelled ``unphysical'' correspond to parameter combinations that are internally inconsistent with the monotonicity of $rA^q_{\rm MBI}(r)$ established in \eqref{rAprime}. Indeed, in Case 2 the function $rA(r)$ is strictly decreasing on $(0,r_1)$ and strictly increasing on $(r_1,\infty)$; consequently its endpoint value
\begin{align}
L_0:=\lim_{r\to0^+}rA^q_{\rm MBI}(r)=-2G(M-E^q_{\rm MBI})
\end{align}
must lie strictly \emph{above} the minimum $r_1A^q_{\rm MBI}(r_1)$, that is $L_0>r_1A^q_{\rm MBI}(r_1)$. Here $M$ is a free parameter, whereas $E^q_{\rm MBI}$ and the minimum $r_1A^q_{\rm MBI}(r_1)$ are both fixed once $q,\beta,G,\kappa_1$ are specified. If the assigned mass satisfies $-2G(M-E)\le r_1A^q_{\rm MBI}(r_1)$, the endpoint value does not lie above the minimum, so the graph of $rA(r)$ cannot satisfy both the boundary behaviour \eqref{rAqMBI0} and the monotonic decrease on $(0,r_1)$ at the same time. Such a mass is not admissible, and the corresponding entry is marked unphysical. The same consistency requirement is understood in Table \ref{horizon} for the general interpolated theories.
\end{remark}

\subsection{Influence of the interpolation and Born parameters}

Since this subsection involves a large amount of numerical analysis, we use the exact but implicit expression of the black hole solution to reduce errors. We also clarify that the numerical analysis below is not intended to cover all results of Section \ref{s4.1}. These results differ only in the values of $M/E^q_{\rm MBI}$ or of
$r_1A(r_1)$, while sharing the same $f(\eta)$. Moreover, a complete scan near the boundary layer ($\kappa_2\to0$) would be repetitive and numerically delicate. We therefore focus on two representative regimes:
the monotonic case $\mathcal{C}\ge0$ (one horizon) and the non-monotonic case $\mathcal{C}<0$ with $r_1A(r_1)<0$ and $M<E$ (two horizons). For each regime, we examine in detail the limiting case $\kappa_1\to1$, in which the interpolated theory approaches the pure Maxwell theory; the horizon radii in this limit are computed with the boundary-layer decomposition \eqref{3.15}, so that the transition from the Born--Infeld to the Maxwell horizon can be traced within the same branch. The remaining branches follow from the same $f(\eta)$ by varying
$M/E^q_{\rm MBI}$ and the sign of $r_1A(r_1)$. Our aim here is to obtain reliable horizon radii in these representative regimes, so as to confirm the classification of Section \ref{s4.1} and to show how the horizons depend on the interpolation and Born parameters. 

Define $f(\eta)=r(\eta)A^q_{\rm MBI}(r(\eta))$, with $r$ related to $\eta$ by
\eqref{3.8}. Then we have
\begin{gather}
f(\eta):=\beta^{\frac14}q^{\frac12}\left(\kappa_1+\frac{\kappa_2}{\sqrt{1-\eta}}\right)^{-\frac12}\eta^{-\frac14}-2G(M-E_{\rm MBI}^q)-\frac{2\pi Gq^{\frac32}}{\beta^{\frac14}}\int^1_{\eta} h(\tau,\kappa_1){\rm d}\tau\label{5.1}\\
\frac{E_{\rm MBI}^{q}}{4\pi}=\frac{q^{\frac32}}{\beta^{\frac14}}\int_{0}^{1}h(\eta,\kappa_1){\rm d}\eta,\label{5.2}
\end{gather}
where $h$ is given by \eqref{3.11}. After setting $q=G = 4\pi=1$, such that \eqref{5.1} reads
\begin{align}
    f(\eta)=\beta^{\frac14}\left(\kappa_1+\frac{\kappa_2}{\sqrt{1-\eta}}\right)^{-\frac12}\eta^{-\frac14}-2\left(M-{\beta^{-\frac14}}\int_{0}^{1}h(\eta,\kappa_1){\rm d}\eta\right)-\frac{1}{2\beta^{\frac14}}\int^1_{\eta} h(\tau,\kappa_1){\rm d}\tau.\label{5.3}
\end{align}
We now give the following plots showing the phenomenon of the event horizons being subject to parameter adjustments. 

\medskip
\textbf{Case 1:} As $\mathcal{C}\ge0$ (hence $\beta\ge4$), we assume $\beta=5$ and $M=2E^q_{\rm MBI}$ to give the following figures with $\kappa_1=0,0.3,0.6,0.9$. 
\begin{figure}[htbp]
  \centering
  \begin{minipage}{0.45\textwidth}
    \centering
    \includegraphics[width=\linewidth]{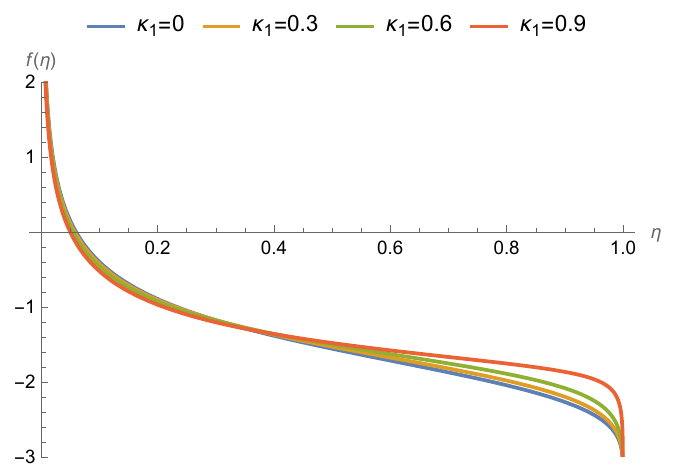}
  \end{minipage}\hfill
  \begin{minipage}{0.45\textwidth}
    \centering
    \includegraphics[width=\linewidth]{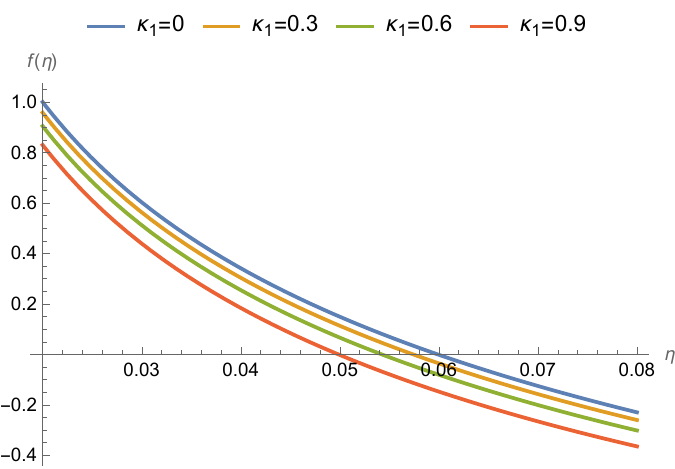}
  \end{minipage}
  \caption{Left: overall curve of $f(\eta)$. Right: magnified view of the intersections with the parametric horizontal axis, showing the event horizon size for different values of $\kappa_1$. Since $\frac{{\rm d}r}{{\rm d}\eta}<0$ given by \eqref{3.13}, the right-hand side figure illustrates the horizon will decrease as $\kappa_1$ decreases.}
  \label{fig:two}
\end{figure}

We also focus on the limiting case $\kappa_1\to1$. As the interpolated theory approximates the pure Maxwell theory extremely, in Section \ref{s3.1} we have concluded that the electric energy is still finite and given by \eqref{3.25b}. Thus, in view of \eqref{3.15} when $\kappa_2$ is less than a certain value (say $\kappa_2\le10^{-9}$), \eqref{5.3} will adopt a new expression ($q=4\pi=1$):
\begin{align}
    f(\eta)=\beta^{\frac14}\left(\kappa_1+\frac{\kappa_2}{\sqrt{1-\eta}}\right)^{-\frac12}\eta^{-\frac14}-2\left(M-\frac4{3\beta^{\frac14}}\right)-\frac{1}{2\beta^{\frac14}}\left(\int^{1-\kappa^2_2\Lambda}_{\eta} h(\tau;\kappa_1)\,{\rm d}\tau+H_2\right).\label{5.4}
\end{align}
Using Table \ref{tab:energylimit}, we give the roots of \eqref{5.4} and values of horizons (by \eqref{3.8})  in the following table.
\begin{table}[htbp]
\centering
\renewcommand{\arraystretch}{1.5}
\begin{tabular}{c|c|c|c|c|c}
\hline
$\kappa_1, \Lambda$            &$1-10^{-9},10^2$ & $1-10^{-12},10^5$ & $1-10^{-15},10^8$ & $1-10^{-18},10^{11}$&$1-10^{-21},10^{14}$\\
\hline
$f(\eta)=0,\eta=$ &0.2560848&0.2520594&0.2519299&0.2519258&0.2519257\\
\hline
Event horizon $r_h^{\rm MBI}=$&1.5544962&1.5606658&1.5608663&1.5608726&1.5608728\\
\hline
\end{tabular}
\caption{Numerical roots of \eqref{5.4} and the corresponding horizon radii. As $\kappa_1$ increases, that is, the Maxwell theory has a dominant weight in \eqref{1.4}, event horizons of the interpolated black hole increase.}
\label{tab:case1horizon}
\end{table}

As shown in Table \ref{tab:case1horizon}, as $\kappa_1$ increases—that is, as the Maxwell term becomes dominant in \eqref{1.4}—the event horizon of the interpolated black hole grows monotonically and approaches the Maxwell value, while remaining distinctly larger than the horizons obtained when the Born--Infeld term carries a non-negligible weight.
\medskip

\textbf{Case 2:} To clearly demonstrate the example of interpolated black holes having two event horizons, we first construct the negative minimum by adjusting parameters. On the one hand, we set $M=0.5E^q_{\rm MBI}$. On the other hand, to satisfy $\mathcal{C}<0$, we fix $4\pi=G=1$, $q=10$ and $\beta=3$. Then  \eqref{5.1} and its derivative now become  
\begin{align}
f'(\eta)=\frac{\beta^{\frac14}}{4\eta^{\frac54}}\left(\kappa_1+\frac{\kappa_2}{\sqrt{1-\eta}}\right)^{-\frac32}
\left[-\left(\kappa_1+\frac{\kappa_2}{\sqrt{1-\eta}}\right)+\frac{\kappa_2\eta}{(1-\eta)^{\frac32}}
\right]+\frac{1}{2\beta^{\frac14}}h(\eta;\kappa_1),\label{fprime}\\
f(\eta)=\beta^{\frac14}\left(\kappa_1+\frac{\kappa_2}{\sqrt{1-\eta}}\right)^{-\frac12}\eta^{-\frac14}+\frac{1}{\beta^\frac14}\int^1_0h(\tau;\kappa_1){\rm d}\eta-\frac{1}{2\beta^{\frac14}}\int^1_{\eta} h(\tau;\kappa_1)\,{\rm d}\tau.
\end{align}
We have the following table to show the negative minimum values with distinct $\kappa_1$.
\begin{table}[htbp]
\centering
\renewcommand{\arraystretch}{1.5}
\begin{tabular}{c|c|c|c|c}
\hline
$\kappa_1$&0&0.3&0.6&0.9\\
\hline
$f'(\eta)=0,\eta_1=$ &0.179313&0.206414 &0.250879&0.356713\\
\hline
$f(\eta_1)=$ &-3.75895&-4.03222&-4.36453&-4.80748\\
\hline
\end{tabular}
\caption{A table of numerical values of the minimum points and values for different $\kappa_1$.}
\label{tab:minimum}
\end{table}

After establishing the conditions that can provide for two event horizons including $\mathcal{C}>0$, $M<E^q_{\rm MBI}$ and $r_1(\eta_1)A(r_1(\eta_1))<0$, we give some figures to show these horizon structures.
\begin{figure}[htbp]
    \centering
    \begin{minipage}[b]{0.4\linewidth}
        \centering
        \includegraphics[width=\linewidth]{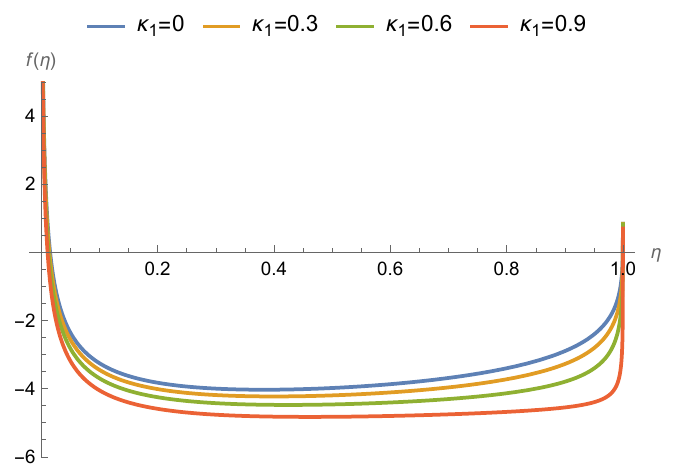}
    \end{minipage}\hfill
    \begin{minipage}[b]{0.4\linewidth}
        \centering
        \includegraphics[width=\linewidth]{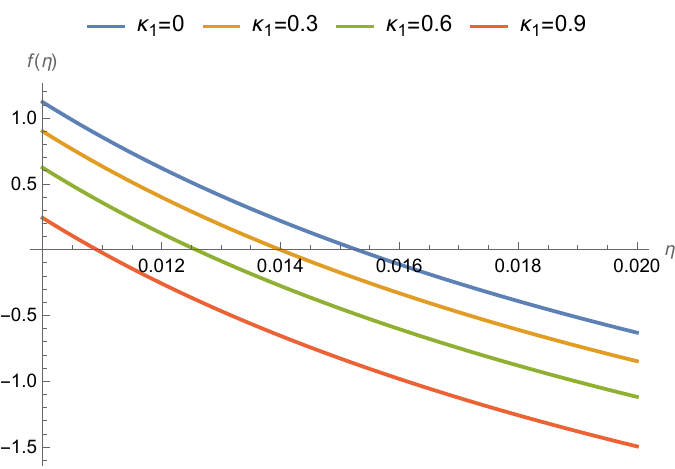}
    \end{minipage}\hfill
    \begin{minipage}[b]{0.4\linewidth}
        \centering
        \includegraphics[width=\linewidth]{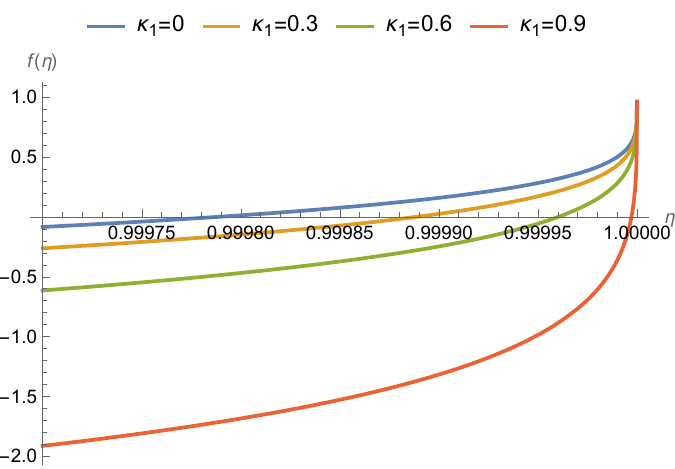}
    \end{minipage}
    \caption{Three plots showing the positions of event horizons.}
    \label{fig:three}
\end{figure}

Figure \ref{fig:three} shows how the event horizons of the interpolated Maxwell--Born--Infeld black hole vary with the interpolation parameter $\kappa_1$. First, as $\kappa_1$ increases---that is, as the weight of the Maxwell term grows and the nonlinear Born--Infeld correction is gradually switched off---the small horizon $r_-$ shrinks while the large horizon $r_+$ grows. By \eqref{3.13}, $\eta$ is inversely proportional to $r$
(equivalently $\mathrm{d}\eta/\mathrm{d}r<0$), so an increase in $\kappa_1$ corresponds to an increase in $r$ at fixed $\eta$. In other words, the larger the Maxwell weight, the more the horizon structure is ``spread out'' in the radial coordinate: the inner horizon moves toward the origin and the outer horizon moves outward, so that the region between the two horizons widens. The physical meaning is that, as the Maxwell term becomes dominant, the defocusing/focusing effect of the electromagnetic field near the origin becomes stronger; the competition between gravity and the electromagnetic force pushes the inner horizon deeper and the outer horizon farther out, so that both the ``dense core'' and the ``outer boundary'' of the black hole are stretched.

Second, in the classical Born--Infeld case $\kappa_1=0$ ($\kappa_2=1$), the small horizon $r_-$ attains its lower bound and the large horizon $r_+$ attains its upper bound; that is, for any $\kappa_1\in[0,1)$,
\begin{align}
r_-(\kappa_1)\ge r_-(0),\quad r_+(\kappa_1)\le r_+(0).
\end{align}
Thus the classical Born--Infeld black hole gives the widest two-horizon configuration among the interpolated black holes: the inner horizon is the innermost and the outer horizon the outermost.

\begin{remark}\label{remark4.1}
The above comparison is carried out on the same branch with $G,M,q,\beta$ fixed, and the interpolated horizons cannot be directly identified with the Reissner--Nordstr\"om horizons $r_\pm^{\rm RN}=GM\pm\sqrt{G^2M^2-4\pi Gq^2}$. The reason is that the Reissner--Nordstr\"om horizons are determined by $G,M,q$ alone, whereas the interpolated horizons depend explicitly on the nonlinear parameter $\beta$: $\beta$ controls the strength of the Born--Infeld correction, and changing $\beta$ shifts the horizon positions.
\end{remark}

\section{Singularity and event horizons of general electric black holes}\label{s5}

In previous sections, we construct the electric black holes arising from the interpolation of Maxwell lagrangian and classical Born--Infeld theory. Then we analyze the singularity and numbers of event horizons of them. However, one of the significant advantages of the interpolation theory framework is that we can not only perturb the pure Maxwell theory by utilizing the nonlinear structure of the classical Born--Infeld theory, but also perturb the pure Maxwell theory by employing the Born--Infeld-type theory with a unique nonlinear structure. Therefore, in this section, we will represent the Lagrange quantity of the interpolation theory as
\begin{align}
    \mathcal{L}(s)=\kappa_1 \mathcal{L}_{\rm M}+\frac{\kappa_2}{\beta}U(2\beta s),\label{MBItype}
\end{align}
and focus on studying the event horizon and black hole singularity of the electric black holes generated by \eqref{MBItype}. Here, the function $U$ may have a truncation threshold, that is, the action $s$ may have an upper bound just like the classical Born--Infeld theory $s<\frac1{2\beta}$. And if $U$ has no truncation threshold, $s=\infty$ is available. Before proceeding to investigations, we recall and assume several physically motivated properties and electrostatic settings of the Lagrangian \eqref{MBItype}.
\begin{enumerate}
    \item The function $U(2\beta s)$ satisfies the standard normalization conditions
    \begin{align}
        U(0)=0,\quad U'(0)=1,\label{normalizedU}
    \end{align}
    and possesses a truncation threshold that enforces $s<\frac1{2\beta}$. This condition, which imposes an upper bound on the electric field strength, embodies the central physical insight introduced by Born and Infeld in their original theory \eqref{1.2}.
    \item As a special case within the Pleba\'nski class of nonlinear electrodynamics \cite{Plebanski1968,Schellstede2016,Russo2024}, when only an electrostatic field is present in $\mathcal{L}_{\rm M}$ and $s$, the function $U(2\beta s)$ satisfies
    \begin{align}
        U'>0,\quad U''>0,\quad '=\frac{\rm d}{{\rm d}s}.\label{plebanski}
    \end{align}
    We emphasize that the weight parameters $\kappa_1,\kappa_2$ within the interpolation framework \eqref{MBItype} do not affect the causal conclusions.
    \item By the Stone--Weierstrass theorem \cite{Stone1948,Yosida1995}, polynomial functions are dense in the space of continuous nonlinear Lagrangians. Consequently, any Born--Infeld type theory can be approximated, in an appropriate topology, by a sequence of polynomial-generated theories:
\begin{align}
    U(\tau)\sim\sum_{n=1}^\infty \zeta_n \tau^n,\quad\zeta_n=\text{constant},\quad\zeta_1=1.\label{stone}
\end{align}
where we have assumed $\zeta_n>0$.
\end{enumerate}

Furthermore, in a pure electrostatic field, the constitutive equation and the Hamiltonian energy density derived from \eqref{MBItype} always take the following forms
\begin{gather}
    \frac{\beta q^2}{r^4}=\left(\kappa_1+2\kappa_2U'(\eta)\right)^2\eta,\quad \eta=2\beta s=\beta({E^r})^2>0,\quad \eta_0>0\label{MBIce}\\
    \mathcal{H}=\frac{1}{\beta}\left[\frac{\kappa_1}{2}\eta+\kappa_2\mathcal{H}_{\rm nonlinear}(\eta)\right],\quad \mathcal{H}_{\rm nonlinear}(\eta)=2\eta U'(\eta)-U(\eta).\label{MBIHamiltonian}
\end{gather}
Accordingly, by \eqref{MBIce}, we have
\begin{gather}
    r={\beta^\frac14}q^\frac12\left(\kappa_1+2\kappa_2U'(\eta)\right)^{-\frac12}\eta^{-\frac{1}{4}},\label{MBIreta}\\
    {\rm d}r=-\frac{\beta^\frac14q^\frac12}{4}(\kappa_1+2\kappa_2U'(\eta))^{-\frac32}\eta^{-\frac54}\left(4\kappa_2\eta U''(\eta)+\kappa_1+2\kappa_2 U'(\eta)\right),\label{MBIdrdeta}\\
    \rho\to\infty,\quad\eta\to0;\quad \rho\to r,\quad \eta\to\eta_0.\label{MBIeta0}
\end{gather}
Next, we consider the electric singularity in framework \eqref{MBItype}.

\subsection{Universal lower bound on the order of the Kretschmann divergence}

In a static spherically symmetric spacetime, the singularity (or Kretschmann curvature) of electric black holes arising from unparametrized Born--Infeld (type) theories always behaves as the Schwarzschild type $K\sim r^{-6}$, and is relegated to $r^{-4}$\cite{Y-book,LiuZhang2025} (or even stronger for an exponential model\cite{yang2022electromagnetic}) when the gravitational mass equals the electromagnetic energy\cite{brewin2007simple}. In \cite{Bronnikov2001}, it is established that there is no regular electric black hole in Born--Infeld (type) theory. Therefore, we now examine whether the singularities of the electric black holes produced by the interpolation theory can be weakened or even regularized.

We simplify the Kretschmann curvature
\begin{align}
K=R_{\mu\nu\alpha\beta}R^{\mu\nu\alpha\beta}=\frac{(r^2{ A}^{\prime\prime})^2+4(r{ A}^\prime)^2+4({ A}-1)^2}{r^4}, \quad A=A(r),\label{formula of Kretschmann 2}
\end{align}
given by the static spherically symmetric spacetime. Note that, although $K$ has been defined in \eqref{formula of Kretschmann}, the present purpose is to yield the singularity of the interpolated Maxwell--Born--Infeld electric black holes. In order to derive the Kretschmann curvature expression for any interpolation model, we do not assume the expression of $A(r)$ here (nor that of $\mathcal{H}$). Inserting \eqref{2.12} and \eqref{2.47} into \eqref{formula of Kretschmann 2}, we have
\begin{align}
K=&\frac{48G^2(M-E)^2}{r^6}+\frac{384\pi G^2(M-E)I}{r^6}+\frac{768\pi^2G^2I^2}{r^6}-\frac{128\pi G^2(M-E)\mathcal H}{r^3}
+\frac{64\pi G^2(M-E)\mathcal H'}{r^2}\notag\\
&-\frac{512\pi^2G^2I\mathcal H}{r^3}
+\frac{256\pi^2G^2I\mathcal H'}{r^2}+64\pi^2G^2r^2(\mathcal H')^2
+256\pi^2G^2\mathcal H^2,\quad \mathcal H'=\frac{{\rm d}\mathcal H}{{\rm d}r},\quad I=\int^r_0\mathcal{H}(\rho)\rho^2{\rm d}\rho.\label{IK}
\end{align}
In view of \eqref{IK}, we first need to consider whether an interpolation theory meeting certain conditions can make $K$ exhibit the Schwarzschild-type behavior. To construct a finite-energy black hole, we note that in the integral $I$ we must have $\mathcal{H}<r^{-3}$ near the origin. In other words, \eqref{MBIHamiltonian} gives
\begin{align}
    \frac{\kappa_1\eta}{2}+\kappa_2(2\eta U'(\eta)-U(\eta))<r^{-3},\quad r\to0,\label{constraint inequality}
\end{align}
which is a constraint inequality that limits the electric field intensity. Moreover, the energy is always finite in the weak field limit $r\to\infty$. To see this, we find from \eqref{plebanski}, \eqref{stone} and \eqref{MBIce} that $\eta\sim r^{-4}\to0$ as $r\to\infty$, so that $\mathcal{H}\sim r^{-4}\to0$ by \eqref{normalizedU}--\eqref{plebanski}. Therefore, the electric energy near infinity, $\int^\infty_r\mathcal{H}(\rho)\rho^2{\rm d}\rho$, is always finite.

\medskip

\textbf{General behaviors of $\eta$, $\mathcal{H}$ and $I$}

In the strong field regime $r\to0$ (hence $\eta\to1^-$), the truncated function $U$, with $\eta\in[0,1)$, satisfies $U'(\eta)\to\infty$. Since $U'$ dominates the constitutive equation \eqref{MBIce} as $\eta\to1^-$ (for $\kappa_2>0$),
\begin{align}
    \frac{\beta q^2}{r^4}\simeq 4\kappa_2^2\,(U')^2
    \quad\Longrightarrow\quad
    U'(\eta)\sim\frac{\sqrt{\beta}\,q}{2\kappa_2\,r^{2}}\to\infty ,
\end{align}
and, by \eqref{MBIHamiltonian} with $2\eta U'-U\simeq 2U'$,
\begin{align}
    \mathcal{H}\sim\frac{2\kappa_2}{\beta}U'
    \sim\frac{q}{\sqrt{\beta}\,r^{2}},\quad r\to0,
\end{align}
which indicates that \eqref{constraint inequality} is valid and that the electric energy is finite for any truncated function $U$ in the interpolation theory. Consequently, $\mathcal{H}r^2$ remains bounded and the integral $I=\int_0^r\mathcal{H}(\rho)\rho^2{\rm d}\rho$ is finite at the origin.

On the other hand, if the function $U$ has no truncation threshold, indicating$\eta\in[0,\infty)$ and  $\eta\to\infty$ as $r\to0$. Then we assume
\begin{align}
    U(\eta)\sim \eta^m\to\infty,\quad m>\frac32.\label{m32}
\end{align}
Inserting \eqref{m32} into \eqref{MBIce}, we have
\begin{align}
    \eta\sim r^{\frac{4}{1-2m}},\quad U'(\eta)\sim r^\frac{4(m-1)}{1-2m},
\end{align}
so that
\begin{align}
    \mathcal{H}r^2\sim r^{\frac{2}{1-2m}}.
\end{align}
Thus, the integral $I$ is finite for the assumption $m>\frac32$ in \eqref{m32}.

Moreover, if $U$ enjoys an exponential growth, we assume that
\begin{align}
    U(\eta)\sim e^{\eta^k}, \quad k>0, \quad\eta\to\infty,\label{Ue}
\end{align}
which, by \eqref{MBIce}, gives
\begin{align}
    \eta\sim\left(\ln\left(\frac1{r^2}\right)\right)^{\frac1k},\quad \mathcal{H}\sim r^{-2}\left(\ln\left(\frac1{r^2}\right)\right)^{\frac{1}{2k}},\quad r\to0.
\end{align}
Since the integral
\begin{align}
    \int\left(\ln\left(\frac1{r^2}\right)\right)^{\frac{1}{2k}}{\rm d}r=-2^{\frac1{2k}}\Gamma\left(1+\frac{1}{2k},\ln\left(\frac1r\right)\right)+C,
\end{align}
is finite for any $k>0$, where $C$ is an integration constant. In particular, for $k=\frac12$ the integrand reduces to unity and the integral is elementary; this is the limiting case of the general formula $-\Gamma(1/(2k),\ln(1/r))+C$, which therefore holds uniformly for all $k>0$. Hence, the exponential-type function $U$ also gives convergent energy.

Consequently, the constraint inequality \eqref{constraint inequality} holds for non-truncated $U$, giving finite electric energy under the premise of \eqref{m32} and \eqref{Ue}.

\medskip
\textbf{Universal bounds on the Kretschmann divergence order in the three cases}

Based on the above results, we now analyze the singularity given by \eqref{formula of Kretschmann 2}. We summarize the asymptotic conclusions for $\mathcal{H}$ and derive the behaviors of $I$ in the following table. We write $L:=\ln\frac1r$.
\begin{table}[htbp]
\centering
\renewcommand{\arraystretch}{2}
\caption{Asymptotic behaviors of $\mathcal{H}$ and $I$ in the strong regime $r\to0$.}
\label{tab7}
\begin{tabular}{c|c|c}
\hline
Type of $U$ growth   & $\mathcal{H}$ as $r\to0$ (Eq.\ \eqref{MBIHamiltonian}) &$I$ as $r\to0$ (Eq.\ \eqref{IK})\\
\hline
Truncated case: $U\to\infty$ as $\eta\to1$& $\mathcal{H}\sim r^{-2}$ &$I\sim {\rm O}(r)$\\
\hline
Polynomial case: $U\sim \eta^m$, $m>\frac32$   & $\mathcal{H}\sim r^{\frac{4m}{1-2m}}$&$I\sim r^{1+\frac{2}{1-2m}}$\\
\hline
Exponential case: $U\sim e^{\eta^k},\quad k>0$ &$\mathcal{H}\sim r^{-2}\bigl(\ln\frac1r\bigr)^{\frac{1}{2k}}$& $I\sim 2^{\frac{1}{2k}}\,r\bigl(\ln\frac1r\bigr)^{\frac{1}{2k}}$\\
\hline
\end{tabular}
\end{table}

In light of Table \ref{tab7}, we can derive the asymptotic behaviors of the Kretschmann curvature for any interpolated theory, as follows.
\begin{enumerate}
    \item \textbf{Truncated case.} Here $\mathcal{H}\sim q/(\sqrt\beta\,r^2)$ and $I\sim qr/\sqrt\beta$. Substituting into \eqref{IK} and collecting the terms surviving at $M=E$, we get
    \begin{align}
        K\sim \frac{48G^2(M-E)^2}{r^6}+\frac{(M-E)C_1}{r^5}+\frac{C_2}{r^4},\label{K1}
    \end{align}
    where $C_1$ and $C_2$ are model-dependent constants depending on $G,\pi,q,\beta$ but not on $\kappa_1,\kappa_2$ (the leading $\mathcal{H}$ is weight-independent). The $r^{-5}$ term carries the factor $(M-E)$ explicitly; it vanishes under $M=E$. Consequently
    \begin{align}
    K\sim
    \begin{cases}
         r^{-6},\quad&\text{Schwarzschild type};\\
        C_2r^{-4},\quad &\text{critical mass-energy condition $M=E$}.
    \end{cases}
    \end{align}
    For the specific interpolated model \eqref{1.4} one finds explicitly $C_2=256\pi^2G^2q^2/\beta$.

    \item \textbf{Polynomial case.} Since $\mathcal{H}\sim r^{4m/(1-2m)}$ and $I\sim r^{1+2/(1-2m)}$, a term-by-term estimate in \eqref{IK} gives
    \begin{align}
        K\sim \frac{48G^2(M-E)^2}{r^6}+\frac{(M-E)C_1}{r^{\frac{3-10m}{1-2m}}}+{C_2}\,r^{\frac{8m}{1-2m}}.\label{K2}
    \end{align}
    The Kretschmann curvature arising from the polynomial case enjoys
    \begin{align}
        K\sim\begin{cases}
            r^{-6},\quad&\text{Schwarzschild type};\\
            C_2r^{\frac{8m}{1-2m}},\quad&\text{critical mass-energy condition $M=E$}.
        \end{cases}
    \end{align}
    Since $m>\frac32$ (which ensures finite electric energy), we have $\frac{8m}{1-2m}\in(-6,-4)$, approaching $-4$ from below as $m\to\infty$. Thus the singularity is relegated to an order no weaker than $r^{-4}$, but it never becomes weaker than $r^{-4}$ for any finite $m$. Note that here both $C_1$ and $C_2$ carry the factor $\kappa_2^{-1/(2m-1)}$ and $\kappa_2^{-2/(2m-1)}$, respectively, so they do depend on the weights; only the exponent is weight-independent.

    \item \textbf{Exponential case.} Since $\mathcal{H}\sim r^{-2}\bigl(\ln\frac1r\bigr)^{\frac1{2k}}$ and $I\sim 2^{\frac1{2k}}r\bigl(\ln\frac1r\bigr)^{\frac1{2k}}$ as $r\to0$, a term-by-term estimate in \eqref{IK} yields
    \begin{align}
        K\sim \frac{48G^2(M-E)^2}{r^6}+\frac{(M-E)C_1}{r^5}\left(\ln\frac1r\right)^{\frac{1}{2k}}+\frac{C_2}{r^4}\left(\ln\frac1r\right)^{\frac{1}{k}}.\label{K3}
    \end{align}
    The Kretschmann curvature arising from the exponential case enjoys
    \begin{align}
        K\sim\begin{cases}
            r^{-6},\quad&\text{Schwarzschild type};\\[1mm]
            C_2r^{-4}\left(\ln\frac1r\right)^{\frac{1}{k}},\quad&\text{critical mass-energy condition $M=E$}.
        \end{cases}\label{K4}
    \end{align}
    In particular, the logarithmic factor never disappears for any $k>0$ (at $k=\tfrac12$ it reads $(\ln\frac1r)^{2}$, not a pure $r^{-4}$); it merely grows more slowly than any power $r^{-\epsilon}$.
\end{enumerate}

Therefore, under $M=E$ the Schwarzschild-type $r^{-6}$ divergence is relegated in all three cases to a weaker singularity whose order is no weaker than $r^{-4}$: precisely $r^{-4}$ in the truncated case, $r^{8m/(1-2m)}\in(-6,-4)$ in the polynomial case, and $r^{-4}(\ln\frac1r)^{1/k}$ in the exponential case. The exponent (order) is independent of $\kappa_1,\kappa_2$, while the coefficient $C_2$ is weight-independent in the truncated and exponential cases (the $\kappa_2$ cancels in the leading $\mathcal{H}$) but weight-dependent in the polynomial case. Hence $r^{-4}$, up to a logarithmic factor, is the weakest admissible singularity under the finite-energy conditions, and a regular electric black hole is excluded in a model-independent way.

We summarize these findings in the following theorem.

\begin{theorem}\label{thm:Ksingularity}
For the general interpolated Maxwell--Born--Infeld type theory \eqref{MBItype} with $\kappa_1+\kappa_2=1$, $\kappa_1\in[0,1)$, $\kappa_2\in(0,1]$, let $U$ satisfy \eqref{normalizedU}--\eqref{stone}, with either a truncation threshold $s<1/(2\beta)$, or, in the non-truncated case, a growth $U\sim\eta^m$ ($m>\frac32$) or $U\sim e^{\eta^k}$ ($k>0$) ensuring finite electric energy. Then the Kretschmann curvature satisfies the asymptotic expressions \eqref{K1}, \eqref{K2} and \eqref{K3} in the truncated, polynomial and exponential cases, respectively. Here $C_1$ and $C_2$ are model-dependent constants independent of $(M-E)$; the $r^{-5}$-type term carries the factor $(M-E)$ and vanishes under $M=E$. The coefficient $C_2$ is independent of $\kappa_1,\kappa_2$ in the truncated and exponential cases, while in the polynomial case it is proportional to $\kappa_2^{-2/(2m-1)}$ and hence weight-dependent. What is universal is therefore the \emph{order} of the leading divergence, not the coefficients. Under the critical mass-energy condition $M=E$ the Schwarzschild-type term $48G^2(M-E)^2r^{-6}$ vanishes in all three cases and the singularity is relegated to an order no weaker than $r^{-4}$: precisely $r^{-4}$ in the truncated case, $r^{8m/(1-2m)}\in(-6,-4)$ in the polynomial case, and $r^{-4}(\ln\frac1r)^{1/k}$ in the exponential case.
\end{theorem}

The key point of Theorem \ref{thm:Ksingularity} is that the asymptotic expressions \eqref{K1}--\eqref{K3}, and hence the relegation of the singularity at $M=E$, have a weight-independent \emph{order}. Indeed, the strong-field behaviors of $\mathcal{H}$ and $I$ collected in Table \ref{tab7} are determined solely by the threshold behavior of $U$ near $\eta\to1^-$ or by its growth as $\eta\to\infty$; the weights enter only through the requirement $\kappa_2>0$. In the truncated and exponential cases even the leading coefficients are weight-independent (the $\kappa_2$ cancels), while in the polynomial case the leading coefficients retain a $\kappa_2^{-1/(2m-1)}$ factor. Consequently, the leading divergence order is the same for all admissible weights, and the conclusion of Theorem \ref{thm:Ksingularity} applies to any Born--Infeld type theory $U$ satisfying the finite-energy conditions stated above. This is consistent with \cite{Bronnikov2001}: even at $M=E$ the $r^{-6}$ divergence is relegated only to $r^{-4}$ (up to a logarithmic factor in the exponential case), so a regular electric black hole remains excluded for any admissible $U$ and any weights.

\begin{remark}\label{remark:tail}
The asymptotic expressions \eqref{K1}--\eqref{K3} retain only the leading singular terms and omit tail terms whose singularities are weaker than the last term displayed. For a specific model, the exact Kretschmann curvature contains infinitely many subleading contributions whose coefficients depend on the detailed structure of $U$ and on $\kappa_1,\kappa_2$; these are of higher order in $r$ and do not affect the leading singularity. In particular, under $M=E$ the singularity of $K$ is dominated by $C_2r^{-4}$ in the truncated case, $C_2r^{8m/(1-2m)}$ in the polynomial case, and $C_2r^{-4}(\ln\frac1r)^{1/k}$ in the exponential case.
\end{remark}

\subsection{Multiplicity event horizons of interpolated electric black holes}

We now turn to establishing a general result of the event horizon for interpolated Maxwell--Born--Infeld type theories of the form \eqref{MBItype} with a truncation threshold $s<1/(2\beta)$, where the nonlinear function $U(2\beta s)$ belongs to a broad class of Born--Infeld type functions. We only consider the truncated case of $U$ in this subsection and the non-truncated case requires a separate treatment and will be commented on at the end of this subsection.

By \eqref{MBIce}-\eqref{MBIeta0}, the total electric energy and the interpolated charged black hole solution read
\begin{align}
\frac{E}{4\pi}=&\int^\infty_0\mathcal{H}(\eta)\rho^2{\rm{d}}\rho=\frac{q^\frac32}{4\beta^\frac14}\int^1_0\left[\frac{\kappa_1}{2}\eta+\kappa_2\mathcal{H}_{\rm nonlinear}(\eta)\right]\cdot\dfrac{2\kappa_2\frac{{\rm d}\mathcal{H}_{\rm nonlinear}(\eta)}{{\rm d}\eta}+\kappa_1}{\eta^\frac74(\kappa_1+2\kappa_2U'(\eta))^\frac52}{\rm d}\eta,\label{MBIE}\\
    A(r)=&1-\frac{2GM}{r}+\frac{8\pi G}{\beta r}\int^\infty_r\left[\frac{\kappa_1}{2}\eta+\kappa_2\mathcal{H}_{\rm nonlinear}(\eta)\right]\rho^2{\rm d}\rho\notag\\
    =&1-\frac{2GM}{r}+\frac{2\pi Gq^\frac32}{\beta^\frac14 r}\int^{\eta_0}_0\left[\frac{\kappa_1}{2}\eta+\kappa_2\mathcal{H}_{\rm nonlinear}(\eta)\right]\cdot\dfrac{2\kappa_2\frac{{\rm d}\mathcal{H}_{\rm nonlinear}(\eta)}{{\rm d}\eta}+\kappa_1}{\eta^\frac74(\kappa_1+2\kappa_2U'(\eta))^\frac52}{\rm d}\eta.\label{MBItypeA}
\end{align}
We now determine the maximum and minimum of $\mathcal{A}(\eta,\kappa_1)$. From \eqref{MBItypeA}, the derivative $(rA(r))'$ is given by
\begin{align}
    \frac{{\rm d}(rA(r))}{{\rm d}r}=&1-8\pi Gr^2\mathcal{H}(r)\notag\\
    =&1-\frac{8\pi Gq}{\sqrt{\beta}}\cdot\frac{\frac{\kappa_1}{2}\eta+\kappa_2\mathcal{H}_{\rm nonlinear}(\eta)}{\sqrt{\eta}\left(\kappa_1+2\kappa_2U'(\eta)\right)}\notag\\
    =&1-\frac{8\pi Gq}{\sqrt{\beta}}\cdot\mathcal{A}(\eta,\kappa_1).\label{generalmathcalA}
\end{align}
To locate the stationary points, we compute the partial derivatives
\begin{align}
     \frac{\partial\mathcal{A}(\eta,\kappa_1)}{\partial\kappa_1}
     &=\frac{U-\eta U'}{\sqrt{\eta}\,(\kappa_1+2\kappa_2U')^2}<0,\label{dAdk1}\\
     \frac{\partial\mathcal{A}(\eta,\kappa_1)}{\partial\eta}
     &=\frac{\left(\frac{\kappa_1\eta}{2}+\kappa_2U\right)
     \left(\kappa_1+2\kappa_2U'+4\kappa_2\eta U''\right)}
     {2\eta^{\frac32}\,(\kappa_1+2\kappa_2U')^2}>0.\label{dAdeta}
\end{align}
Indeed, by \eqref{normalizedU}--\eqref{plebanski} we have $U(0)=0$, $U'>0$, $U''>0$; the convexity of $U$ gives $U<\eta U'$ for $\eta>0$, whence $\frac{\partial\mathcal{A}}{\partial\kappa_1}<0$, while $\frac{\kappa_1\eta}{2}+\kappa_2U>0$, $\kappa_1+2\kappa_2U'>0$ and $4\kappa_2\eta U''>0$, whence $\frac{\partial\mathcal{A}}{\partial\eta}>0$. Thus $\mathcal A$ is strictly decreasing in $\kappa_1$ and strictly increasing in $\eta$, and it admits no stationary points in the interior of $(\eta,\kappa_1)\in(0,1)\times[0,1)$; its extreme values are attained only on the boundary $\eta\to0^+$ and $\eta\to1^-$:
\begin{align}
    \lim_{\eta\to1^-}\mathcal{A}(\eta,\kappa_1)
    =\lim_{\eta\to1^-}\left(1-\frac{U(\eta)}{2\eta\,U'(\eta)}\right)=1,
    \quad
    \lim_{\eta\to0^+}\mathcal{A}(\eta,\kappa_1)=0,
\end{align}
for every fixed $\kappa_2\in(0,1]$, where $U'/U\to\infty$ near the truncation threshold and $U\sim\eta$ near $\eta=0$. In light of \eqref{generalmathcalA}, we conclude that
\begin{gather}
    0\le\mathcal{A}(\eta,\kappa_1)<1,
    \\[2mm]
    1-\frac{8\pi Gq}{\sqrt{\beta}}<(rA(r))'\le 1.
\end{gather}
Thus, the monotonicity of $rA(r)$ is controlled solely by gravitational constant $G$, the electric charge $q$, and the parameter $\beta$. 

We now analyze the monotonicity of $rA(r)$. On the one hand, if $\sqrt{\beta}\ge{8\pi Gq}$, then $rA(r)$ is monotonically increasing for all $r>0$. Using \eqref{MBIE}--\eqref{MBItypeA}, we find
\begin{align}
    (rA(r))_{\rm minimum}=&\lim_{r\to0,\eta_0\to1,\kappa_1=0}[rA(r)]\notag\\
    =&\lim_{r\to0,\eta_0\to1}\left(r-2GM+\frac{2\pi Gq^\frac32}{\beta^\frac14}\int^{\eta_0}_0\frac{\mathcal{H}_{\rm nonlinear}(\eta)\cdot\frac{{\rm d}\mathcal{H}_{\rm nonlinear}(\eta)}{{\rm d}\eta}}{2^\frac32\eta^\frac74(U'(\eta))^\frac52}{\rm d}\eta\right)\notag\\
    =&2G(E-M).
\end{align}
It is evident that the relation between the black hole mass and the electromagnetic energy governs the existence of event horizons: 
\begin{enumerate}
    \item If $E\ge M$, no event horizon exists and the singularity at $r=0$ is naked; this scenario is ruled out by the Cosmic Censorship Hypothesis.
    \item If $E< M$, there exists precisely one event horizon.
\end{enumerate}

On the other hand, if $\sqrt{\beta}<{8\pi Gq}$, then $rA(r)$ possesses a unique minimum at some point $r_1>0$, corresponding to a value $\eta_1\in(0,1)$. It is essential to determine the value of $r_1$ (or $\eta_1$). From the $rr$-component of the Einstein equations, we have
\begin{align}
    \frac{{\rm d}(rA(r))}{{\rm d}r}\Big|_{r=r_1}=0\iff 1=8\pi Gr_1^2\mathcal{H}(\eta_1).
\end{align}
Using \eqref{MBIHamiltonian} and \eqref{MBIreta}, this condition can be rewritten as
\begin{align}
    \frac{8\pi Gq}{\sqrt{\beta}}=\frac{(\kappa_1+2\kappa_2 U'(\eta_1))\eta_1^{\frac12}}{\frac{\kappa_1}{2}\eta_1+2\kappa_2\eta_1 U'(\eta_1)-\kappa_2U(\eta_1)}:=N(\eta_1;\kappa_1,\kappa_2),\quad \eta_1\in(0,1).\label{MBIr1eta1}
\end{align}
Clearly, $N(\eta_1;\kappa_1,\kappa_2)$ is non-negative, and for any admissible $\kappa_1$ and $\eta_1$, it satisfies
\begin{align}
\frac{\partial N}{\partial\kappa_1}&=\frac{\eta_1^{\frac12}(\eta_1U'(\eta_1)-U(\eta_1))}{\left[\mathcal{H}_{\rm nonlinear}(\eta_1)+\kappa_1\left(\frac{\eta_1}{2}-\mathcal{H}_{\rm nonlinear}(\eta_1)\right)\right]^2}\ge0,\\
    \frac{\partial N}{\partial\eta_1}&=\frac{-\frac{\kappa_1\eta_1}{4}\left(\kappa_1+2\kappa_2U'(\eta_1)\right)-\frac{\kappa_2U(\eta_1)}{2}\left(\kappa_1+2\kappa_2U'(\eta_1)\right)-\kappa_2\eta_1 U''(\eta_1)\left(\kappa_1\eta_1+2\kappa_2U(\eta_1)\right)}{\eta_1^{\frac12}\left(\frac{\kappa_1}{2}\eta_1+\kappa_2\mathcal{H}_{\rm nonlinear}(\eta_1)\right)^2}\le0,
\end{align}
Since ${8\pi Gq}>\sqrt{\beta}$, equation \eqref{MBIr1eta1} admits a unique solution $\eta_1$ for any combination of $\kappa_1$ and $\kappa_2$, provided that $N(\eta_1;\kappa_1,\kappa_2)>1$, i.e.,
\begin{align}
    {(\kappa_1+2\kappa_2 U'(\eta_1))\eta_1^{\frac12}}>{\frac{\kappa_1}{2}\eta_1+2\kappa_2\eta_1 U'(\eta_1)-\kappa_2U(\eta_1)}.\label{MBINScondition}
\end{align}
We now verify that $N(\eta_1;\kappa_1,\kappa_2)>1$ holds for arbitrary $U$ and all $(\kappa_1,\eta_1)\in[0,1)\times(0,1)$. From the monotonicity of $N$ with respect to $\kappa_1$ and $\eta_1$, we deduce
\begin{align}
    N(\eta_1;\kappa_1,\kappa_2)\in(1,\infty),\label{MBINrange}
\end{align}
which is consistent with \eqref{MBINScondition} and guarantees that \eqref{MBIr1eta1} always possesses a unique solution $r_1$ (equivalently $\eta_1$). 

We summarize these findings in the following theorem.
\begin{theorem}\label{thm1}
For any interpolated Maxwell--Born--Infeld type theory $\mathcal{L}(s)$ with a truncation threshold $s<\frac1{2\beta}$, in a radially symmetric electrostatic field (so that $s=\frac12(E^r)^2$), the following statements hold.
\begin{enumerate}
    \item If the function $rA(r)$ corresponding to the charged black hole solution \eqref{MBItypeA} satisfies
\begin{align}
    0\le1-\frac{8\pi Gq}{\sqrt{\beta}}<(rA(r))'\le 1,\quad '=\frac{\rm d}{{\rm d}r},\label{thm1.1}
\end{align}
then its minimum value is determined by $G$ and the black hole mass $M$ and energy $E$, namely $rA(r)=2G(E-M)$, which directly dictates the existence of the event horizon:
\begin{align}
M>E, \text{ one event horizon;}\quad M\le E, \text{ no event horizon}.\label{EMrelation1}
\end{align}
In particular, a naked singularity occurs when $M=E$.
\item If \eqref{thm1.1} is replaced by 
\begin{align}
        1-\frac{8\pi Gq}{\sqrt{\beta}}<(rA(r))'\le 1,\quad 1-\frac{8\pi Gq}{\sqrt{\beta}}<0,\label{EMrelation2}
\end{align}
then $rA(r)$ always possesses a minimum point, which is obtained by solving \eqref{MBIr1eta1}, since \eqref{MBINScondition} and \eqref{MBINrange} are always satisfied for some $\eta\in(0,1)$ and $\kappa_1\in[0,1)$.
\end{enumerate}
\end{theorem}

We append several remarks concerning Theorem \ref{thm1}.
\begin{remark}\label{remark1}
(i) In view of \eqref{MBIE}, we emphasize that this theorem provides theoretical guidance for numerical investigations of event horizon existence: the numerical values of the charge $q$, the field strength parameter $\beta$, and the black hole mass $M$ must be chosen in accordance with \eqref{EMrelation1}. 

(ii) In the second case of Theorem \ref{thm1}, although the existence of a minimum is guaranteed, the sign of the minimum value $r_1A(r_1)$ cannot be determined without specifying the explicit form of $U(\eta)$, since $r_1$ (or $\eta_1$) is obtained from \eqref{MBIr1eta1}.

\end{remark}

Motivated by the second observation in Remark \ref{remark1}, we now enumerate the possible horizon configurations, which depend on the value of $r_1A(r_1)$. It is important to note that the asymptotic behavior of $A(r)$ as $r\to0$ and $r\to\infty$ also influences the horizon structure, making it necessary to examine the finiteness of the electric energy. To this end, we use \eqref{MBIce} to rewrite \eqref{MBIHamiltonian} as
\begin{align}
    \mathcal{H}=\sqrt{\frac{\eta(D^r)^2}{\beta}}-\frac{\kappa_1\eta}{2\beta}-\frac{\kappa_2 U(\eta)}\beta,\quad D^r=\frac{q}{r^2},\quad U(\eta)>0.\label{MBIHamiltonian2}
\end{align}
On the one hand, as $r\to\infty$ and $\eta\to0$ (hence $U\to0$), we have
\begin{align}
    \mathcal{H}\le\sqrt{\frac{\eta(D^r)^2}{\beta}}<{\rm O}(r^{-2}),
\end{align}
which implies that the energy term in \eqref{2.45} is finite and
\begin{align}
    rA(r)\to\infty \text{ as }r\to\infty.\label{MBIAinfty}
\end{align}
On the other hand, as $r\to0$, using \eqref{stone} and $\kappa_1+\kappa_2=1$, we assume the asymptotic behavior
\begin{align}
    \kappa_1s+\frac{\kappa_2}\beta U(2\beta s)\sim s+\kappa_2s^n,\quad n\ge2,
\end{align}
from which \eqref{MBIce} yields
\begin{align}
    \eta\sim r^{-\frac{4}{2n-1}}.
\end{align}
Substituting this into \eqref{MBIHamiltonian2}, we obtain
\begin{align}
    r^2\mathcal{H}\sim {\rm O}\left(r^{-\frac{2}{2n-1}}\right),\quad \frac{2}{2n-1}\le\frac23<1,
\end{align}
which ensures that the energy term in \eqref{2.47} is finite and that
\begin{align}
    rA(r)\sim{-2G(M-E)}\begin{cases}
        >0,\quad&M<E;\\
        <0,\quad &M>E;\\
        =0,\quad &M=E,
    \end{cases}\quad r\to0.\label{MBIAzero}
\end{align}
We are now in a position to state the following theorem, which provides a complete classification of event horizons.
\begin{theorem}
    For any interpolated Maxwell--Born--Infeld type theory $\mathcal{L}(s)$ with a truncation threshold $s<\frac1{2\beta}$ in a radially symmetric electrostatic field (so that $s=\frac12(E^r)^2$), taking into account \eqref{EMrelation2}, \eqref{MBIAinfty}, and \eqref{MBIAzero}, the existence of event horizons is classified as follows:
\begin{table}[htbp]
\centering
\caption{Horizon classification based on the sign of $r_1A(r_1)$ and mass-energy conditions.}
\renewcommand{\arraystretch}{1.5}
\begin{tabular}{c|c|c}
\hline
    Sign of $r_1A(r_1)$ & Mass-energy condition & Number of horizons \\
\hline
$=0$ & $M < E$ & 1 at $r_1$ \\
$=0$ & $M \ge E$ & 0 (unphysical case)  \\
\hline
$<0$ & $M < E$ & 2 \\
$<0$ & $M = E$ & 1  \\
$<0$ & $M > E,\; r_1A(r_1) < -2G(M-E)$ & 1  \\
$<0$ & $M > E,\; r_1A(r_1) \ge -2G(M-E)$ & 0 (unphysical case) \\
\hline
$>0$ & $M\ge E$ & 0 (unphysical case) \\
$>0$ & $M<E,\; r_1A(r_1) < -2G(M-E)$& 0 (naked singularity)\\
$>0$ & $M<E,\; r_1A(r_1) \ge -2G(M-E)$& 0 (unphysical case)\\
\hline
\end{tabular}
\label{horizon}
\end{table}
\end{theorem}

It is worth comparing the above classification with the multi-horizon solutions known in non-truncated nonlinear electrodynamics. In \cite{Gao2021}, black holes with three and four horizons were constructed for Lagrangians of the form $\mathcal{L}=\sum_{i\ge1}\alpha_i(F^2)^i$, which possess no truncation threshold; the horizon structure was obtained by prescribing the roots of $A(r)$ and reconstructing $\mathcal{L}$ through the Lagrangian-generation method, and the resulting Lagrangian depends only on the coupling constants $\alpha_i$, not on $M$ or $q$. Earlier multi-horizon and regular black hole solutions in nonlinear electrodynamics were reported in \cite{NojiriOdintsov2017}. The present analysis differs in two respects: first, it is restricted to the truncated Born--Infeld type class $s<1/(2\beta)$, for which $\mathcal A(\eta,\kappa_1)<1$ provides a uniform lower bound $\mathcal C=1-8\pi Gq/\sqrt{\beta}$ for $(rA)'$ and yields the model-independent classification of Table \ref{horizon}; second, it proceeds in the opposite direction, starting from a given interpolated Lagrangian and deducing the number of horizons, rather than reconstructing the Lagrangian from a prescribed horizon structure. The latter route has been followed, for instance, in \cite{GaoLuYuShen2018}, where multi-horizon and multi-singularity solutions were obtained by prescribing the metric and reconstructing the Lagrangian; the Lagrangian so obtained is tailored to the chosen metric and does not belong to the Born--Infeld type, in contrast to the interpolated Lagrangians considered here.

\section{Conclusions and comments}\label{s6}

In this paper, we have systematically investigated electrically and magnetically charged black hole solutions within an interpolated Maxwell--Born--Infeld theory, whose Lagrangian is a convex combination of the Maxwell and classical Born--Infeld Lagrangians \eqref{1.4}. This framework provides a controlled interpolation between linear and nonlinear electrodynamics, governed by the single weight parameter $\kappa_1$ (equivalently $\kappa_2$). We now summarize how the questions raised in the Introduction are answered.

\medskip

\textbf{Boundedness of the electric energy in the Maxwell limit.}
The electric field remains finite at the origin for all $\kappa_1<1$, guaranteeing a finite electromagnetic self-energy. As the theory approaches the pure Maxwell theory, the total energy remains bounded: the value $1/2$ given by \cite{LY1} and obtained by interchanging the limit and the integration in the normalization energy function \eqref{3.12}, is a mathematical artifact of non-uniform convergence, while the correct sharp upper bound is $4/3$, as established by the boundary-layer decomposition \eqref{3.15} together with the limits \eqref{boundary layer} and \eqref{regular}, and confirmed numerically in Tables \ref{tab:Lambda} and \ref{tab:energylimit}; the resulting energy--radius relation is \eqref{3.25b}. The additional contribution arises from an infinitesimally thin boundary layer near the origin, whose resolution requires splitting the energy integral into a regular part and a boundary-layer part rather than a direct application of Fatou's lemma in\cite{LY1}.

\medskip

\textbf{Interpolation of the black hole and behavior of the singularity.}
In the electrostatic case we have constructed explicit black hole solutions and analyzed their asymptotic behavior near the origin and at infinity, \eqref{3.21}--\eqref{3.22}. The Hamiltonian energy density is monotonic in $\kappa_1$ \eqref{3.27}, so that the metric interpolates monotonically between the classical Born--Infeld and Reissner--Nordstr\"om solutions, \eqref{3.25}; the electric field saturates at the Born--Infeld maximal value even when the Born--Infeld weight is arbitrarily small \eqref{3.29}. The Kretschmann curvature exhibits a robust singularity structure: it is of the Schwarzschild type $r^{-6}$ in the generic case, and under the critical condition $M=E$ the leading term is cancelled so that the singularity is relegated to $r^{-4}$ \eqref{Kretschmann of electric} or $r^{-4}(\ln \frac1r)^{\frac{1}k}$ \eqref{K4}, independently of the interpolation parameters. Furthermore, the horizon structure is governed solely by $\mathcal{C}=1-8\pi Gq/\sqrt{\beta}$ \eqref{rAprime}: when $\mathcal{C}\ge0$ the existence of horizons is decided entirely by the mass--energy relation, whereas when $\mathcal{C}<0$ the function $rA(r)$ possesses a unique minimum and the horizon configurations are classified in Table \ref{tab:MBIhorizon}, with the explicit dependence on the interpolation and Born parameters illustrated in Tables \ref{tab:case1horizon} and \ref{tab:minimum} and Figures \ref{fig:two}--\ref{fig:three}.

\medskip

\textbf{Status of magnetic and dyonic configurations.}
Magnetostatic and dyonic black hole solutions have also been constructed and their asymptotic expansions derived, with the magnetic metric expressed in terms of the Gauss hypergeometric function \eqref{3.31} and the dyonic metric given by \eqref{3.32}. Consistent with the flat-spacetime results of \cite{LY1}, the point-particle energies diverge at the origin, ruling out magnetic monopoles and dyons on energetic grounds. In this instance the energy conditions supply no independent obstruction: as established in Section \ref{s3}, the weak, dominant, and strong energy conditions are all satisfied in the electric, magnetic, and dyonic sectors alike. The inadmissibility of magnetic and dyonic configurations is therefore encoded entirely in the divergence of their total energy rather than in any violation of the energy conditions. The energy conditions thus play a diagnostic role: their continued validity confirms that the interpolated theory is free of the exotic-matter pathologies that would otherwise obscure the purely energetic obstruction against magnetic and dyonic configurations.

\medskip

\textbf{General interpolated Born--Infeld-type theories.}
Beyond the specific interpolated model, we have extended the analysis to general interpolated Lagrangians \eqref{MBItype} with a nonlinear structure $U$ satisfying the normalization conditions \eqref{normalizedU}, the causal conditions \eqref{plebanski} of the Pleba\'nski class, and the polynomial approximation \eqref{stone}. For this class, Theorem \ref{thm:Ksingularity} shows that the Kretschmann curvature always exhibits the Schwarzschild-type $r^{-6}$ divergence near the origin in the generic case, and that under $M=E$ the relegation stops at the universal lower bound $r^{-4}$, namely $K\sim C_2r^{-4}$ in the truncated case \eqref{K1}, $K\sim C_2r^{8m/(1-2m)}$ with $8m/(1-2m)\in(-6,-4)$ for $m>3/2$ in the polynomial case \eqref{K2}, and $K\sim C_2r^{-4}(\ln\frac1r)^{1/k}$ with $k>0$ in the exponential case \eqref{K3}. Since the lower bounds do not involve $\kappa_1,\kappa_2$ (although the precise exponent may depend on the growth class of $U$ and the coefficient $C_2$ may depend on $\beta$), the conclusion applies to arbitrary Born--Infeld-type theories and yields a model-independent exclusion of regular electric black holes. For the event horizons, Theorem \ref{thm1} and Table \ref{horizon} provide a complete classification for general interpolated theories with a truncation threshold: the black holes may possess zero, one, or two event horizons according to the mass-energy relation and the sign of $r_1A(r_1)$ at the unique minimum $r_1$, and the parameters governing the transitions between these configurations are identified.

\medskip

In summary, the interpolated Maxwell--Born--Infeld (type) theory provides a consistent and physically rich bridge between the Maxwell and Born--Infeld (type) limits. Its distinctive features--a finite electric energy with the sharp upper bound $4/3$ in the Maxwell limit, a monotonic interpolation of the black-hole metric, a model-independent lower bound on the order of the Kretschmann singularity, and a complete classification of horizon configurations--are all absent in either limiting theory. The electromagnetic asymmetry of the model is realized entirely at the level of energy: all classical energy conditions remain satisfied, so the theory is free of exotic-matter pathologies, and yet magnetic and dyonic point charges are excluded by their divergent self-energy, an obstruction that no energy-condition veto is required to enforce.

\end{document}